\documentclass[12pt]{article}
\usepackage[T1]{fontenc}
\usepackage[utf8]{inputenc}
\usepackage{geometry}
\usepackage{amsmath}
\usepackage{amssymb}
\usepackage{graphicx}

\makeatletter
\pdfoutput=1 
\usepackage{graphicx}
\usepackage{amsfonts,amsthm,amsmath,amssymb,upgreek}
\usepackage{hyperref}
\usepackage{cite}
\usepackage{units}
\usepackage{color}
\usepackage[export]{adjustbox}
\usepackage{mathtools}
\usepackage{physics}

\usepackage{mciteplus}
\numberwithin{equation}{section}

\mciteSetMidEndSepPunct{}{\ifmciteBstWouldAddEndPunct.\else\fi}{\relax}
\renewcommand\[{\begin{equation}} 
\renewcommand\]{\end{equation}}
\renewenvironment{align*}{\align}{\endalign}

\renewcommand{\title}[1]{\vbox{\center\bf{\Large{#1}}}\vspace{5mm}}
\renewcommand{\author}[1]{\vbox{\center#1}\vspace{5mm}}
\newcommand{\address}[1]{\vbox{\center\em#1}}
\newcommand{\email}[1]{\vbox{\center\tt#1}\vspace{5mm}}

\makeatother

\begin{document}
\begin{titlepage}

\begin{center} 
\hfill \\ 
\hfill \\ 
\vskip .5cm

\title{Quantum Epidemiology: Operator Growth, Thermal Effects, and SYK}

\author{Xiao-Liang Qi$^a$  and Alexandre Streicher$^{ab}$} 

\address{$^{a}$ Stanford Institute for Theoretical Physics, Stanford University,\\ Stanford, CA 94305, USA

\vspace{10pt}

$^{b}$ Department of Physics, University of California,\\ Santa Barbara, CA 93106, USA
}

\email{xlqi@stanford.edu | alex@physics.ucsb.edu} 

\end{center} 
\begin{abstract}
In many-body chaotic systems, the size of an operator generically
grows in Heisenberg evolution, which can be measured by certain out-of-time-ordered
four-point functions. However, these only provide a coarse probe of
the full underlying operator growth structure. In this article we
develop a methodology to derive the full growth structure of fermionic
systems, that also naturally introduces the effect of finite temperature.
We then apply our methodology to the SYK model, which features all-to-all
$q$-body interactions. We derive the full operator growth structure
in the large $q$ limit at all temperatures. We see that its temperature
dependence has a remarkably simple form consistent with the slowing
down of scrambling as temperature is decreased. Furthermore, our finite-temperature
scrambling results can be modeled by a modified epidemic model, where
the thermal state serves as a vaccinated population, thereby slowing
the overall rate of infection.
\end{abstract}
\end{titlepage}

\tableofcontents{}

\section{Introduction \& Summary}

In chaotic quantum many-body systems, operators grow in size as time
evolves. For example, in spatially local systems one expects that
the extent of an operator $\mathcal{O}\left(t\right)$ grows as 
\begin{equation}
\frac{d}{dt}\textrm{Volume}\left[\mathcal{O}\left(t\right)\right]\propto\textrm{Surface Area}\left[\mathcal{O}\left(t\right)\right]\label{localgrowth}
\end{equation}
since the new terms generated by taking $\left[H,\mathcal{O}\left(t\right)\right]$
will live on the boundary of the domain of $\mathcal{O}\left(t\right)$
\cite{Lieb:1972wy,hastings2010locality,Roberts:2014isa,Roberts:2016wdl}.
Consequently, the extent grows linearly with an effective \textquotedbl speed
of light\textquotedbl{} $\simeq vt$. Up to exponential error all operators
outside the effective light-cone will commute with $\mathcal{O}\left(t\right)$.
This effective speed of light is known as the Lieb-Robinson velocity
\cite{Lieb:1972wy}. This highlights the fact that space can be a
derived concept in quantum mechanics, as without the Hamiltonian there
may not be a sense in which one piece of the Hilbert space factorization
is closer to another.

Now we would like to contrast this behavior with that exhibited by
$q$-local systems, where the Hamiltonian couples all the degrees
of freedom together in $q$-body interactions. Consequently, there
is no notion of spatial locality, and we accordingly refer to such
interactions as coupling together ``internal'' degrees of freedom.
Yet, there remains structure in the evolution of operators in these
systems, as we often find that the sizes of operators grow exponentially
\begin{equation}
\frac{d}{dt}\mathrm{Size}\left[\mathcal{O}\left(t\right)\right]\propto\mathrm{Size}\left[\mathcal{O}\left(t\right)\right]\label{qlocalgrowth}
\end{equation}
where by size we mean the number of simple operators multiplied together
in a typical piece of $\mathcal{O}\left(t\right)$. The intuition
behind this growth is that the percentage of the $q$-body interactions
utilized in $\left[H,\mathcal{O}\left(t\right)\right]$ is proportional
to the size of $\mathcal{O}\left(t\right)$, and almost all the resultant
operators obtained from $\left[H,\mathcal{O}\left(t\right)\right]$
are bigger than $\mathcal{O}\left(t\right)$ \cite{dankert2009exact,Sekino:2008he,brown2010convergence,Brown:2012gy,Bentsen:2018uph,2016PhRvE..93d2138H,Roberts:2018mnp}.\footnote{Eq. (\ref{qlocalgrowth}) actually can be considered as the same formula
as in Eq. (\ref{localgrowth}) applied to a completely connected graph,
so that the area of a region is proportional to its volume \cite{Bentsen:2018uph}.}

Systems with both spatial locality and a large number of internal
degrees of freedom--such as (chaotic) field theories in the large-$N$
limit -- display both linear spatial growth and exponential internal
size growth \cite{Shenker:2014cwa,Nahum:2017yvy,Xu:2018xfz,Xu:2018dfp,Chen:2018hjf}.
The growth of an evolving simple operator $W\left(t\right)$ can be
probed using another simple operator $V$, using the (anti-)commutator
squared $\ev*{\left|\left[W\left(t\right),V\right]\right|^{2}}$ or
their corresponding out-of-time-order correlator (OTOC) $\left\langle W^{\dagger}\left(t\right)V^{\dagger}W\left(t\right)V\right\rangle $
\cite{Larkin:1969abc,Shenker:2013pqa,Shenker:2013yza,Kitaev:2014t1,kitaev2018soft,Shenker:2014cwa,Aleiner:2016eni,Gu:2016oyy,stanford2016many,Patel:2017vfp}.

In order to develop the coarse-grained profile of operator growth,
one must compute many OTOCs. The ``chaos bound'' \cite{Maldacena:2015waa}
obeyed by OTOCs suggests that after an initial dissipation time, they
de-correlate no faster than exponentially, with a rate $\lambda_{L}$
no larger than $2\pi T$ where $T=1/\beta$ is the temperature. This
implies that presence of the thermal state $\rho\propto\exp\left(-\beta H\right)$
slows down the effective growth rate of operators as temperature is
decreased.

The Heisenberg evolution of operator $\mathcal{O}\left(t\right)$
is independent of temperature, so the entire effect of temperature
must be contained in the matrix elements of $\mathcal{O}$. Therefore,
the natural finite temperature generalization of operator size has
remained an open question (one recent proposal is given in \cite{lucas2018operator}). 

In this article, we address this issue by characterizing not only
the average size of an operator but its entire size distribution.
We then can define the effective size distribution of an operator
at finite temperature by how it \textit{changes} the size of the square
root of thermal density operator $\rho^{1/2}$ (we explain why this
is a natural choice in section (\ref{sec:Thermal-Operators})). 

This definition leads to some nontrivial general results, independent
of the details of the specific physical system. In particular, we
observe that in generic fermion systems, the effective size of a single
fermion operator is ``thermally renormalized'' to a value $\delta_{\beta}=G\left(\beta/2\right)$
smaller than $1$, where $G\left(\tau\right)$ is the thermal two-point
function. The size of the thermal operator $\rho^{1/2}$ itself is
$\frac{N}{2}\left(1-\delta_{\beta}\right)$, determined by the same
renormalization factor $\delta_{\beta}$. To gain a more explicit
understanding, we will work in the context of the SYK model \cite{Sachdev:1992fk,Kitaev:2014t2},
a $q$-local Hamiltonian built out of $N$ flavors of Majorana fermions,
which saturates the chaos bound at low temperatures \cite{Maldacena:2016hyu,Kitaev:2014t1,Kitaev:2014t2}

The remainder of the article is organized as follows. We begin in
section (\ref{sec:Operator-Distributions-and}) by building up the
notion of operator ``size''. First, we show that one may expand
any operator $\mathcal{O}\left(t\right)$ in an orthonormal operator
basis of the unique products of Majorana flavors. In the doubled theory,
the operator basis maps to an orthonormal basis of states in the doubled
Hilbert space. We define a ``size'' operator $n$ in the doubled
theory counting the average number of flavors in an operator basis
state. We are then able to demonstrate that four-point functions measure
the average ``size'' of an operator. Therefore, the de-correlation
of a thermal OTOC is exactly equivalent to the growing average size
of the operator $\psi_{1}\left(t\right)\rho^{1/2}$ 
\[
-\frac{1}{N}\sum_{j=1}^{N}\Tr\left(\rho^{1/2}\psi_{1}\left(t\right)\psi_{j}\psi_{1}\left(t\right)\rho^{1/2}\psi_{j}\right)=1-\frac{2}{N}n\left[\psi_{1}\left(t\right)\rho^{1/2}\right]
\]
The average size of the operator $\psi_{1}\left(t\right)\rho^{1/2}$
starts at $n\left[\psi_{1}\rho^{1/2}\right]\approx n\left[\rho^{1/2}\right]=\frac{N}{2}\left(1-\delta_{\beta}\right)$
and then grows sigmoidally in time, eventually saturating (scrambling)
at a value of $N/2$.

Up to this point we have been discussing the average size of the operator
$\psi\left(t\right)\rho^{1/2}$. In fact, the entire size distribution
of this operator has physical significance. Hence, in section (\ref{sec:Thermal-Operators})
we construct generating functions for operator size distributions
by inserting a weighting factor $\exp\left(-\mu n\right)$. First,
we study the size distribution of the thermal operator $\rho^{1/2}$
by setting up a generating function $\mathcal{Z}_{\mu}\left[\rho^{1/2}\right]$,
which is similar to a grand canonical partition function. Next, we
show that the fractional distance to scrambling for the operator $\rho^{1/2}$
is always given by $\delta_{\beta}\equiv1-n/n_{*}=G\left(\beta/2\right)<1$.
Then, we set up the generating function for the size distribution
of $\psi\left(t\right)\rho^{1/2}$, which we find naturally splits
into a product of $\mathcal{Z}_{\mu}\left[\rho^{1/2}\right]$ and
a modified two-point function $\mathcal{G}_{\mu}\left(t\right)$.
We show that the $\mu$-expansion of $\mathcal{G}_{\mu}\left(t\right)$
determines the growth distribution induced by multiplying $\rho^{1/2}$
by $\psi\left(t\right)$, and that $\mathcal{G}_{\mu}\left(t\right)$
is simply the two-point function for the original theory with a $\mu$-dependent
twisted boundary condition. We conclude the section by noting that
on average, the size increase induced by a single fermion is given
by the fractional scrambling distance $\delta_{\beta}$, which leads
us to propose that $\delta_{\beta}$ should be interpreted as a thermally
renormalized unit of size.

Everything in sections (\ref{sec:Operator-Distributions-and}) and
(\ref{sec:Thermal-Operators}) applies to general fermionic systems.
In sections (\ref{sec:SYK-Model}) and (\ref{sec:SYK-Operator-Growth}),
we apply this methodology to the large-$q$ SYK model. Solving the
large-$N$ saddle point equation in the large $q$ limit with our
$\mu$-dependent twist, we obtain the full operator growth structure.
After a dynamical renormalization of coupling constant $\mathcal{J}$
(which after a short amount of time essentially amounts to replacing
the coupling with a smaller $\beta$-dependent constant) and a renormalization
of size unit from $1$ to $\delta_{\beta}$, we observe that the full
growth structure has the same functional form as the infinite temperature
case. The dynamical renormalization of the coupling is the signature
of the slowdown of the effective growth rate as temperature is decreased. 

We conclude the section by discussing how to understand this finite
temperature slowdown of scrambling in an epidemic model, where the
thermal factor effectively vaccinates a large subset of the population,
thereby slowing down the overall infection rate. We end the paper
by discussing implications and future directions in section (\ref{sec:discussion}).

\section{Operator Distributions and Two-Sided Wavefunctions\label{sec:Operator-Distributions-and}}

As an operator $\mathcal{O}\left(t\right)$ evolves in time, it becomes
supported along operators of increasing size. This can be inferred
from the Heisenberg equation of motion $\dot{\mathcal{O}}\left(t\right)=i\left[H,\mathcal{O}\left(t\right)\right]$.
Now, in order to properly discuss how much one operator is supported
along another, we need an operator inner product. When the Hilbert
space is finite-dimensional it is natural to use the Frobenius inner
product: $\braket*{\mathcal{O}_{A}}{\mathcal{O}_{B}}\equiv\mathrm{Tr}(\mathcal{O}_{A}^{\dagger}\mathcal{O}_{B})$.
We may then expand operators in an orthonormal operator basis, which
amounts to inserting a complete set of operators $\left\{ \Gamma_{I}\right\} $
\[
\mathcal{O}\left(t\right)=\sum_{I}\Gamma_{I}\mathrm{Tr}\left(\Gamma_{I}^{\dagger}\mathcal{O}\left(t\right)\right)\equiv\sum_{I}c_{I}\left(t\right)\Gamma_{I}
\]
Note that at this point we have set up a Hilbert space of operators.
If the original Hilbert space $\mathbb{H}$ has dimension $L$, the
operator Hilbert space is $\mathbb{H}\otimes\overline{\mathbb{H}}$,
with dimension $L^{2}$.

\subsection{Purification \label{subsec:Purification}}

\begin{figure}[tb]
\begin{centering}
\includegraphics[width=1\textwidth]{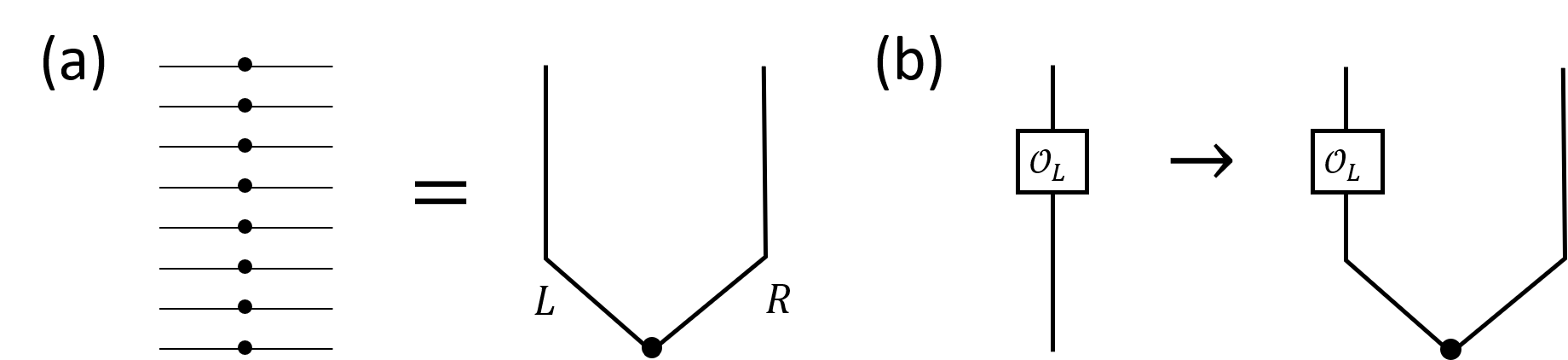} 
\par\end{centering}
\caption{Illustration of the purification procedure that maps operators to
states in a doubled Hilbert space. (a) A maximally entangled state
$|0\rangle$ (Eq. (\ref{eq:state0}) which can be viewed as many EPR
pairs between the two systems. (b) The mapping between operator $\mathcal{O}$
and the corresponding state $|\mathcal{O}\rangle$ obtained by applying
$\mathcal{O}$ to the left system (see Eq. (\ref{eq:state for operator O})).
\label{fig:purification}}
\end{figure}
Since $\mathbb{H}\otimes\overline{\mathbb{H}}$ is isomorphic to $\mathbb{H}\otimes\mathbb{H}$,
one can always maps each operator to a quantum state in a ``doubled''
system with Hilbert space dimension $L^{2}$. More explicitly, this
mapping is defined by considering two copies of the original physical
system, named as $L$ and $R$, and introducing a maximally entangled
state $|0\rangle$ (see Fig. \ref{fig:purification}(a)). For any
maximally entangled state, there is a basis choice of the form $\left\{ \ket{n}_{L}\otimes\ket{m}_{R}\right\} $
such that 
\begin{equation}
\ket{0}=\sum_{m,n}\delta_{mn}\ket{n}_{L}\otimes\ket{m}_{R}=\sum_{n}\ket{n}_{L}\otimes\ket{n}_{R}\label{eq:state0}
\end{equation}
For later convenience we have chosen the norm of the state to be $\braket{0}=L$.

Then the operator-to-state mapping is defined by 
\begin{equation}
\mathcal{O}\rightarrow|\mathcal{O}\rangle\equiv\mathcal{O}_{L}\otimes\mathbb{I}_{R}|0\rangle\label{eq:state for operator O}
\end{equation}
where $\mathcal{O}_{L}$ is the operator $\mathcal{O}$ acting on
the Hilbert space of the left system, as is illustrated in Fig. \ref{fig:purification}(b).
It is easy to verify that the inner product $_{LR}\braket{\mathcal{O}_{A}}{\mathcal{O}_{B}}_{LR}=\mathrm{Tr}(\mathcal{O}_{A}^{\dagger}\mathcal{O}_{B})$
is determined by the Frobenius inner product of the corresponding
operators. Our orthonormal basis of operators $\left\{ \Gamma_{I}\right\} $
will thereby serve as an orthonormal basis of states $\ket{\Gamma_{I}}$.
Thus, the problem of understanding how $\mathcal{O}\left(t\right)$
is distributed across a particular choice of basis operators is equivalent
to understanding how the two-sided state $\ket{\mathcal{O}\left(t\right)}$
is distributed across a particular choice of two-sided basis states.
Since the choice of maximally entangled state $|0\rangle$ is not
unique, the operator-to-state mapping has an ambiguity of $U(L)$
acting on the $R$ system. Since the same transformation is performed
to $|\mathcal{O}\rangle$ and the basis vector $|\Gamma_{I}\rangle$,
all our discussion will be independent from this freedom of basis
choice.

\subsubsection{Orthonormal Basis of Operators\label{subsec:Orthonormal Basis of Operators}}

One of the simplest algebras with an interesting finite dimensional
representation is the algebra of $N$ flavors of Majorana fermions,
where $N$ is even: 
\begin{equation}
\left\{ \psi_{i},\psi_{j}\right\} =2\delta_{ij}\label{eq:Majorana Algebra}
\end{equation}
Note that this implies that $\psi_{j}^{2}=1$, which will be convenient
for our purposes, unlike the more common convention where $\left\{ \psi_{i},\psi_{j}\right\} =\delta_{ij}$
and thus $\psi_{j}=1/2$. Such operators are traceless, Hermitian,
and unitary. Furthermore, the algebra is invariant under taking any
single $\psi_{i}\rightarrow-\psi_{i}$, and so the product of any
subset of the $N$ fermions is also traceless. Thus, it is easy to
construct an orthogonal operator basis by taking unique ordered products
of Majorana fermions 
\begin{equation}
\Gamma_{I}\equiv\Gamma_{i_{1}i_{2}...i_{k}}=i^{\frac{k\left(k-1\right)}{2}}\psi_{i_{1}}...\psi_{i_{k}}\qquad1\leq i_{1}<i_{2}<...<i_{k}\leq N\label{eq:majoranastring}
\end{equation}
where the pre-factor has been inserted so that the resultant $\Gamma_{I}$
matrices are Hermitian. All nontrivial $\Gamma_{I}$ (with $k>0$)
are traceless. Since the product $\Gamma_{I}\Gamma_{J}$ is also a
string of fermions, which is only trivial when $I=J$ (when the two
strings are identical and Majorana fermions pairwise cancel), we have
\begin{equation}
\mathrm{Tr}\left(\Gamma_{I}^{\dagger}\Gamma_{J}\right)=\mathrm{Tr}\left(\Gamma_{I}\Gamma_{J}\right)=\delta_{IJ}\Tr\left(1\right)\label{eq:orthonormal basis operators}
\end{equation}
Furthermore, the basis operators $\Gamma_{I}$ have simple algebraic
relations, since they either commute or anti-commute according to
the relation 
\[
\Gamma_{I}\Gamma_{J}=\left(-1\right)^{\left|I\right|\left|J\right|+\left|I\cap J\right|}\Gamma_{J}\Gamma_{I}
\]
where $\left|I\right|$ is the number of elements in the multi-index
$I$.

\subsubsection{Mapping Basis Operators to Basis States}

\begin{figure}[tb]
\begin{centering}
\includegraphics[width=1\textwidth]{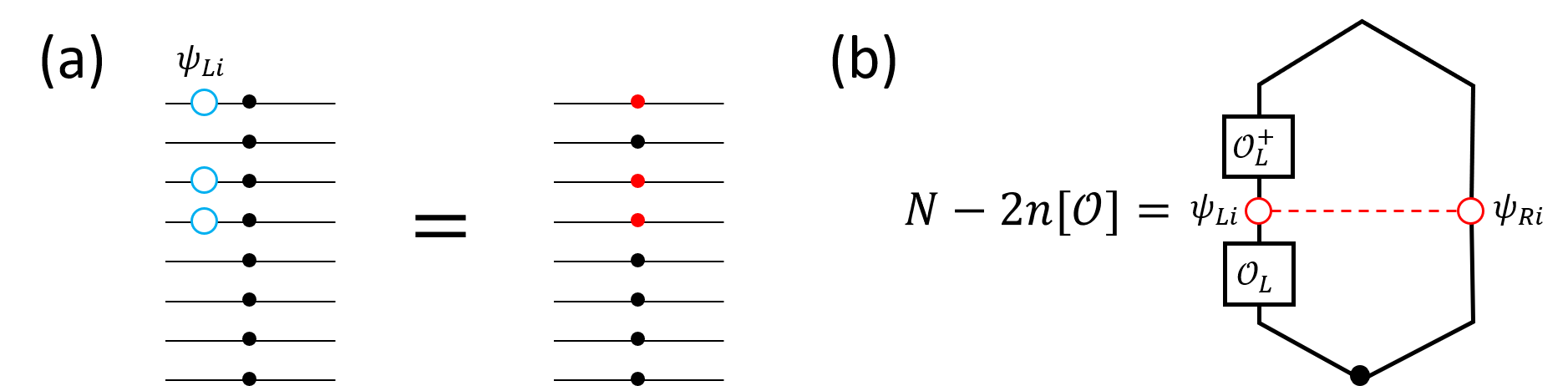} 
\par\end{centering}
\caption{\label{fig:purification2} (a) The mapping of a Majorana string $\Gamma_{I}$
in Eq. (\ref{eq:majoranastring}) to a state in the doubled system.
Each fermion operator $\psi_{Li}$ creates a fermion (red dot) while
the fermions that are absent in $\Gamma_{I}$ stays in the vacuum
state with fermion number $0$ (black dot). (b) Illustration of the
relation between average size of operator $\mathcal{O}$ and OTOC. }
\end{figure}

The purification isomorphism is quite simple to realize, as has been
discussed by \cite{Maldacena:2018lmt,Gu:2017njx}. We consider two
copies of the original system, which contains $2N$ Majorana fermions
labeled by $\psi_{j}^{L}$ and $\psi_{j}^{R}$, $j=1,2,...,N$. We
then define a maximally entangled state $\ket{0}$, 
\begin{equation}
\left(\psi_{j}^{L}+i\psi_{j}^{R}\right)\ket{0}=0,\;\forall j\label{eq:Definition of 0}
\end{equation}
We may think of this state as a vacuum (all spins down, all bits set
to 0) with regards to a set of entangled complex fermions operators
\[
c_{j}\ket{0}=0\qquad c_{j}\equiv\frac{\psi_{j}^{L}+i\psi_{j}^{R}}{2}\quad\left\{ c_{j},c_{k}\right\} =\left\{ c_{j}^{\dagger},c_{k}^{\dagger}\right\} =0\quad\left\{ c_{j},c_{k}^{\dagger}\right\} =\delta_{jk}
\]
where $c_{j}^{\dagger}=\left(c_{j}\right)^{\dagger}$. Since state
$|0\rangle$ is the ground state of a quadratic Hamiltonian $H=\sum_{j}c_{j}^{\dagger}c_{j}$,
it is straightforward to compute the entanglement entropy\cite{2003JPhA...36L.205P}
and verify that the state is maximally entangled between $L$ and
$R$. As we discussed earlier, the choice of $|0\rangle$ is not unique,
but this choice is convenient for our purpose. The basis operators
$\Gamma_{I}$ are mapped to states in the doubled system of $2N$
Majorana fermions: 
\begin{equation}
\ket{\Gamma_{I}}\equiv\Gamma_{I}^{L}\ket{0}=i^{\frac{k\left(k-1\right)}{2}}\psi_{i_{1}}^{L}...\psi_{i_{k}}^{L}\ket{0}=i^{\frac{k\left(k-1\right)}{2}}c_{i_{1}}^{\dagger}...c_{i_{k}}^{\dagger}\ket{0}=c_{i_{k}}^{\dagger}...c_{i_{1}}^{\dagger}\ket{0}\label{eq:orthonormal basis states}
\end{equation}
Therefore each basis operator $\Gamma_{I}$ is mapped to a particular
fermion configuration in the doubled system, with fermions $i_{1},i_{2},...,i_{k}$,
as is illustrated in Fig. \ref{fig:purification2}(a). Essentially,
the identity operator maps to the vacuum and nontrivial operators
are mapped to excitations in the doubled theory.

\subsection{Four-Point Functions Probe Operator Size \label{subsec:Four-Point-Function-and}}

At this point, we can discuss the number operator $n_{j}\equiv c_{j}^{\dagger}c_{j}$,
which returns $1$ when applied to basis states containing the flavor
$j$ and zero otherwise 
\begin{align}
n_{j} & \equiv c_{j}^{\dagger}c_{j}=\frac{1}{2}\left(1+i\psi_{j}^{L}\psi_{j}^{R}\right) & \matrixel{\Gamma_{I}}{n_{j}}{\Gamma_{J}} & =\delta_{j\in J}\braket{\Gamma_{I}}{\Gamma_{J}}\label{eq:nj operator}
\end{align}
Thus, we see that for a generic operator $\mathcal{O}$, the expectation
value of $n_{j}$ returns the percentage of basis operators in $\mathcal{O}$
containing flavor $j$. Furthermore, we note that this expectation
value is closely related to a one-sided four-point function (see Fig.
\ref{fig:purification2}), since 
\begin{align}
\ev{\left(2n_{j}-1\right)}{\mathcal{O}} & =\ev{i\psi_{j}^{L}\psi_{j}^{R}}{\mathcal{O}}=\ev{\left(\mathcal{O}^{L}\right)^{\dagger}\psi_{j}^{L}i\psi_{j}^{R}\mathcal{O}^{L}}{0}=-\ev{\left(\mathcal{O}^{L}\right)^{\dagger}\psi_{j}^{L}\mathcal{O}^{L}i\psi_{j}^{R}}{0}\nonumber \\
 & =\ev{\left(\mathcal{O}^{L}\right)^{\dagger}\psi_{j}^{L}\mathcal{O}^{L}\psi_{j}^{L}}{0}=\mathrm{Tr}_{L}\left(\left(\mathcal{O}^{L}\right)^{\dagger}\psi_{j}^{L}\mathcal{O}^{L}\psi_{j}^{L}\right)\nonumber \\
\Rightarrow\ev{\left(2n_{j}-1\right)}{\mathcal{O}} & =\mathrm{Tr}\left(\mathcal{O}^{\dagger}\psi_{j}\mathcal{O}\psi_{j}\right)\label{eq:Flavor Number Operator to Four Point Function}
\end{align}
Here we have assumed $\mathcal{O}$ to be fermionic. In the first
two steps, we simply plugged in the definitions of $n_{j}$ and $\ket{\mathcal{O}}$.
In the third step, we anti-commuted $i\psi_{j}^{R}$ through $\mathcal{O}^{L}$,
as right fermionic operators anti-commute with left fermionic operators.
Then, we used the definition of $\ket{0}$ (\ref{eq:Definition of 0})
to replace $-i\psi_{j}^{R}\ket{0}$ with $\psi_{j}^{L}\ket{0}$. Afterwards,
we had an expectation value of only left operators for a maximally
entangled state, so we traced out the right Hilbert space entirely,
leaving us with an infinite temperature four-point function of the
left-only system.

The relationship between operator quantities and one-sided correlators
is simpler in terms of the anti-commutator squared, since we have
\begin{align}
\frac{1}{4}\Tr\left(\left\{ \mathcal{O},\psi_{j}\right\} ^{\dagger}\left\{ \mathcal{O},\psi_{j}\right\} \right) & =\frac{1}{2}\Tr\left(\mathcal{O}^{\dagger}\mathcal{O}\right)+\frac{1}{2}\Tr\left(\mathcal{O}^{\dagger}\psi_{j}\mathcal{O}\psi_{j}\right)=\frac{1}{2}\braket{\mathcal{O}}+\frac{1}{2}\ev{\left(2n_{j}-1\right)}{\mathcal{O}}\nonumber \\
\Rightarrow\frac{1}{4}\Tr\left(\left\{ \mathcal{O},\psi_{j}\right\} ^{\dagger}\left\{ \mathcal{O},\psi_{j}\right\} \right) & =\ev{n_{j}}{\mathcal{O}}\equiv n_{j}\left[\mathcal{O}\right]\label{eq:Flavor Number Operator to Anti-Commutator}
\end{align}
where we used (\ref{eq:Flavor Number Operator to Four Point Function})
to replace $\Tr\left(\mathcal{O}^{\dagger}\psi_{j}\mathcal{O}\psi_{j}\right)$
with $\left(2n_{j}-1\right)$. One should note that if $\mathcal{O}$
is bosonic, the right-hand side of (\ref{eq:Flavor Number Operator to Four Point Function})
will acquire a minus sign and the anti-commutators in (\ref{eq:Flavor Number Operator to Anti-Commutator})
will be replaced with commutators. We denote the average value of
$n_{j}$ in operator $\mathcal{O}$ as $n_{j}[\mathcal{O}]$.

We can also define a total number operator (a.k.a. size operator)
that returns the number of flavors or size of a basis state 
\begin{align}
n & \equiv\sum_{j=1}^{N}n_{j}=\sum_{j=1}^{N}c_{j}^{\dagger}c_{j} & \matrixel{\Gamma_{I}}{n}{\Gamma_{J}} & =\left|I\right|\braket{\Gamma_{I}}{\Gamma_{J}}\label{eq:size operator}
\end{align}
with $|I|$ the number of Majorana fermion operators in the string
$\Gamma_{I}$. Consequently, $\ev{n}{\mathcal{O}}$ is the average
number of flavors in the operator $\mathcal{O}$, or the average size
of the operator $\mathcal{O}$. By flavor averaging Eq. (\ref{eq:Flavor Number Operator to Anti-Commutator}),
we see that the flavor-averaged anti-commutator squared measures the
average size of the operator $\mathcal{O}$ 
\[
\frac{1}{4N}\sum_{j=1}^{N}\Tr\left(\left\{ \mathcal{O},\psi_{j}\right\} ^{\dagger}\left\{ \mathcal{O},\psi_{j}\right\} \right)=\frac{\ev{n}{\mathcal{O}}}{N}\equiv\frac{n\left[\mathcal{O}\right]}{N}
\]
where the anti-commutators are replaced with commutators if $\mathcal{O}$
is bosonic.

Alternatively, we may flavor average Eq. (\ref{eq:Flavor Number Operator to Four Point Function})
in order to relate the flavor-averaged four-point function to the
average size 
\begin{equation}
\frac{\left(-1\right)^{\left|\mathcal{O}\right|}}{N}\sum_{j=1}^{N}\Tr\left(\mathcal{O}^{\dagger}\psi_{j}\mathcal{O}\psi_{j}\right)=\ev{\left(1-\frac{2n}{N}\right)}{\mathcal{O}}\equiv\ev{\delta}{\mathcal{O}}\equiv\delta\left[\mathcal{O}\right]\label{eq:Four-Point Function to Size}
\end{equation}
where $\left(-1\right)^{\left|\mathcal{O}\right|}$ is 1 if $\mathcal{O}$
is bosonic and $-1$ if $\mathcal{O}$ is fermionic. 

Noting that the number of unique products of $k$ Majoranas Eq. (\ref{eq:majoranastring})
goes as $\binom{N}{k}$, we see that the most common size is that
of $N/2$. Indeed, a generic operator should be equally supported
across all unique products, leading to a distribution of the form
$P_{k}=\binom{N}{k}/2^{N}$. Thus, a totally scrambled operator has
a size $n_{*}=N/2$, so we see that the flavor-averaged four-point
function measures the average fractional distance $1-2n/N$ an operator's
size is from this scrambled value. Therefore we define the fractional
scrambling distance operator 
\begin{equation}
\delta\equiv1-\frac{n}{n_{*}}=1-\frac{2n}{N}\label{eq:FracScramDisOp}
\end{equation}

\subsection{Operator Size Generating Function}

\label{sec:generatingfunc}

By defining the number operator $n$, we can now go beyond the average
operator size probed by four-point functions. Rather than just the
average, we study all moments systematically by introducing a generating
function\cite{2016PhRvE..93d2138H} 
\[
\mathcal{Z}_{\mu}[\mathcal{O}]=\langle\mathcal{O}|e^{-\mu n}|\mathcal{O}\rangle
\]
By taking derivatives of the generating function we can obtain all
moments of $n$: 
\[
\langle\mathcal{O}|n^{k}|\mathcal{O}\rangle=\frac{(-1)^{k}}{k!}\left.\frac{\partial^{k}\mathcal{Z}}{\partial\mu^{k}}\right|_{\mu=0}
\]
A more useful expansion is a Taylor expansion in $e^{-\mu}$: 
\begin{equation}
\mathcal{Z}_{\mu}[\mathcal{O}]=\sum_{n=0}^{N}e^{-\mu n}P_{n}[\mathcal{O}]\label{eq:expansion}
\end{equation}
in which the coefficients $P_{n}\left[\mathcal{O}\right]$ is the
percentage of terms in $\mathcal{O}$ having size $n$.

\section{Including Temperature\label{sec:Thermal-Operators}}

The main goal of the current work is to understand the role of temperature
in operator growth. After all, the dynamics of the operator $\psi_{1}\left(t\right)$
under Heisenberg evolution has no knowledge about temperature.

One natural way to introduce temperature is to consider the operator
$\rho^{1/2}$ where $\rho=Z_{\beta}^{-1}\exp\left(-\beta H\right)$
is thermal state at inverse temperature $\beta$. The purification
of $\rho^{1/2}$ is the thermofield double (TFD) state $\ket{TFD}$
(the other factor of $\rho^{1/2}$ from the full $\rho$ is used to
make $\bra{TFD}$) 
\[
|TFD\rangle=Z_{\beta}^{-1/2}e^{-\frac{\beta}{4}\left(H_{L}+H_{R}\right)}|0\rangle=Z_{\beta}^{-1/2}e^{-\frac{\beta}{2}H_{L}}|0\rangle=\left|\rho^{1/2}\right\rangle 
\]
where the Hamiltonians $H_{L},H_{R}$ are required to satisfy the
condition $(H_{L}-H_{R})|0\rangle=0$. This state is a natural choice
for studying thermodynamic properties, because for each operator $\mathcal{O}$,
we can consider the corresponding operator $\mathcal{O}\rho^{1/2}$,
and its average size will be directly measured by the finite temperature
four-point function: 
\begin{equation}
\delta\left[\mathcal{O}\rho^{1/2}\right]=1-\frac{n\left[\mathcal{O}\rho^{1/2}\right]}{N/2}=\frac{(-1)^{|\mathcal{O}|}}{N}\sum_{j=1}^{N}\Tr\left(\rho^{1/2}\mathcal{O}^{\dagger}\psi_{j}\mathcal{O}\rho^{1/2}\psi_{j}\right)\label{eq:4ptfiniteT}
\end{equation}

Now, we are interested in uncovering the size distribution of a time
evolved ``thermal'' operator $\mathcal{O}\left(t\right)\rho^{1/2}$.
To do so, we will define a generating function for its size moments
$\mathcal{Z}_{\mu}\left[\mathcal{O}\left(t\right)\rho^{1/2}\right]$.
We shall find that such a generating function naturally factorizes
into a product of the generating function for the Gibbs state $\mathcal{Z}_{\mu}\left[\rho^{1/2}\right]$
and a ``connected'' piece $\mathcal{G}_{\mu}\left[\mathcal{O}\left(t\right)\right]$.
By using this to extract the size distribution of $\rho^{1/2}$ from
the size distribution of $\mathcal{O}\left(t\right)\rho^{1/2}$, we
find a natural definition for a ``thermal'' size of $\mathcal{O}\left(t\right)$.

\subsection{Thermal State}

\label{subsec:thermal state}

We begin by studying $\rho^{1/2}$. Taking $\mathcal{O}=\rho^{1/2}$
in Eq. (\ref{eq:Four-Point Function to Size}), we find the following
relation between the thermal two-point function and the size operator
\begin{align*}
G\left(\frac{\beta}{2}\right)=\frac{1}{N}\sum_{j=1}^{N}Z_{\beta}^{-1}\Tr\left(e^{-\beta H}\psi_{j}\left(\frac{\beta}{2}\right)\psi_{j}\right) & =1-\frac{n\left[\rho^{1/2}\right]}{N/2}=\delta\left[\rho^{1/2}\right]
\end{align*}
This relation tells us that the most de-correlated value of the Euclidean
two-point function - $G\left(\beta/2\right)$ - is equal to the fractional
distance the operator $\rho^{1/2}$ is from being scrambled 
\begin{equation}
\delta_{\beta}\equiv\delta\left[\rho^{1/2}\right]=G\left(\frac{\beta}{2}\right)\label{eq:Average Thermal State Distance Two-Point Function}
\end{equation}

which implies that the average size of $\rho^{1/2}$ is given by 
\begin{equation}
n\left[\rho^{1/2}\right]=\frac{N}{2}\left(1-G\left(\frac{\beta}{2}\right)\right)\label{eq:Average Thermal State Size Two-Point Function}
\end{equation}

In the high temperature limit $\beta\rightarrow0$, one expects $G(\beta/2)\simeq G(0)=1$,
since the fermions square to one (\ref{eq:Majorana Algebra}), which
is consistent with the fact that $\rho^{1/2}$ approaches identity
and the size shrinks to zero. On the contrary, in the low temperature
limit $\beta\rightarrow\infty$, if $G(\beta/2)\rightarrow0$, the
size of $\rho^{1/2}$ approaches the scrambled (typical) value $N/2$.
This result is very general since the imaginary time two-point function
$G(\tau)$ decays in most physical systems. For example, in all systems
with a unique ground state and an excitation gap, $G(\tau)$ decays
exponentially at low temperature limit, so that $G(\beta/2)\rightarrow0$
when $\beta\rightarrow\infty$. In a conformal field theory, $G(\tau)$
decays in power law in the zero temperature limit, which also leads
to the same length $n_{\beta\rightarrow\infty}=N/2$.

To learn more than just the average size, we construct the generating
function 
\begin{equation}
\mathcal{Z}_{\mu}\left[\rho^{1/2}\right]=\left\langle \rho^{1/2}\right|e^{-\mu n}\left|\rho^{1/2}\right\rangle =\ev{e^{-\mu n}}{TFD}\label{eq:thermalZmuTFD}
\end{equation}
Therefore learning about the operator distribution of $\rho^{1/2}$
is equivalent to learning about the fermion number distribution in
the thermofield double state.

\subsection{Thermal Fermion\label{subsec:Thermal-Fermion}}

If we take $\mathcal{O}=\psi_{1}\left(t\right)$ in Eq. (\ref{eq:4ptfiniteT}),
we see that the average size of $\psi\left(t\right)\rho^{1/2}$ is
entirely equivalent to an out-of-time-order correlator (a.k.a. OTOC)
\cite{Larkin:1969abc,Shenker:2013pqa,Shenker:2013yza,Kitaev:2014t1}:
\[
-\frac{1}{N}\sum_{j=1}^{N}\Tr\left(\rho^{1/2}\psi_{1}\left(t\right)\psi_{j}\psi_{1}\left(t\right)\rho^{1/2}\psi_{j}\right)=1-\frac{2}{N}n\left[\psi_{1}\left(t\right)\rho^{1/2}\right]\equiv\delta\left[\psi_{1}\left(t\right)\rho^{1/2}\right]
\]
Therefore, the statement that the OTOC de-correlates exponentially
is equivalent to the statement that the average size of the operator
$\psi_{1}\left(t\right)\rho^{1/2}$ grows exponentially. If the OTOC
vanishes in long time, that implies that the size of $\psi_{1}(t)\rho^{1/2}$
reaches the scrambled value $n_{*}=N/2$.

The size distribution of $\psi_{1}(t)\rho^{1/2}$ can be uncovered
through the generating function 
\begin{equation}
\mathcal{Z}_{\mu}\left[\psi_{1}(t)\rho^{1/2}\right]=\ev*{e^{-\mu n}}{\psi_{1}\left(t\right)\rho^{1/2}}=\ev*{\psi_{1}^{L}(t)e^{-\mu n}\psi_{1}^{L}(t)}{TFD}\label{eq:thermalfermionZmu}
\end{equation}
The operator $e^{-\mu n}$ can be viewed as an Euclidean time evolution
with time $\mu$ and Hamiltonian $n$, so that the generating function
(\ref{eq:thermalfermionZmu}) is related to the two-point function
in a system with time-dependent Euclidean evolution: 
\begin{equation}
\mathcal{G}_{\mu}(\tau_{a},\tau_{b})=\frac{\left\langle 0\right|\mathcal{T}\left[e^{-\beta\left(H_{L}+H_{R}\right)/2}e^{-\mu n\left(\beta/4\right)}\psi_{1}^{L}(\tau_{a})\psi_{1}^{L}(\tau_{b})\right]\left|0\right\rangle }{\left\langle 0\right|\mathcal{T}\left[e^{-\beta\left(H_{L}+H_{R}\right)/2}e^{-\mu n\left(\beta/4\right)}\right]\left|0\right\rangle }\label{eq:twopointfunc}
\end{equation}
where $\mathcal{T}$ is the Euclidean time ordering symbol and $\psi_{1}^{L}(\tau_{a,b})$
are imaginary time evolved fermion operators. Note that the denominator
in (\ref{eq:twopointfunc}) is, up to a factor of thermal partition
function $Z_{\beta}$ that cancels with the numerator, exactly the
size generating function $\mathcal{Z}_{\mu}\left[\rho^{1/2}\right]$
in Eq. (\ref{eq:thermalZmuTFD}). Therefore, the size generating function
$\mathcal{Z}_{\mu}\left[\psi_{1}(t)\rho^{1/2}\right]$ naturally factorizes
into the product of 
\begin{equation}
\mathcal{Z}_{\mu}\left[\psi_{1}(t)\rho^{1/2}\right]=\mathcal{G}_{\mu}\left(\frac{\beta^{+}}{4}+it,\frac{\beta^{-}}{4}+it\right)\mathcal{Z}_{\mu}\left[\rho^{1/2}\right]\label{eq:Near-Gibbs Generator}
\end{equation}

This equation clarifies that the two-point function $\mathcal{G}_{\mu}\left(\beta/4^{+}+it,\beta/4^{-}+it\right)$
measures the size change $\psi_{1}(t)$ induces upon $\rho^{1/2}$
through multiplication. We can see this directly by applying the expansion
in Eq. (\ref{eq:expansion}) to both sides of this equation, in order
to obtain the following convolution formula for the size distribution
of $\psi_{1}\left(t\right)\rho^{1/2}$

\begin{equation}
P_{n}\left[\psi_{1}(t)\rho^{1/2}\right]=\left(K^{\beta}\left[\psi_{1}\left(t\right)\right]*P\left[\rho^{1/2}\right]\right)_{n}=\sum_{m=0}^{n}K_{m}^{\beta}\left[\psi_{1}(t)\right]P_{n-m}\left[\rho^{1/2}\right]\label{eq:convolution equation}
\end{equation}
with $K_{m}^{\beta}\left[\psi_{1}(t)\right]$ defined by the expansion
of $\mathcal{G}_{\mu}\left(\beta/4^{+}+it,\beta/4^{-}+it\right)$
in powers of $e^{-\mu}$: 
\begin{equation}
\mathcal{G}_{\mu}\left(\frac{\beta^{+}}{4}+it,\frac{\beta^{-}}{4}+it\right)=\sum_{m=0}^{N}e^{-m\mu}K_{m}^{\beta}\left[\psi_{1}(t)\right]\label{eq:moments of G}
\end{equation}
In this sense, $K_{m}^{\beta}$ can be viewed as the ``growth distribution''
caused by applying $\psi_{1}(t)$ to the thermal state $\rho^{1/2}$.

\begin{figure}[tb]
\begin{centering}
\includegraphics[width=0.5\textwidth]{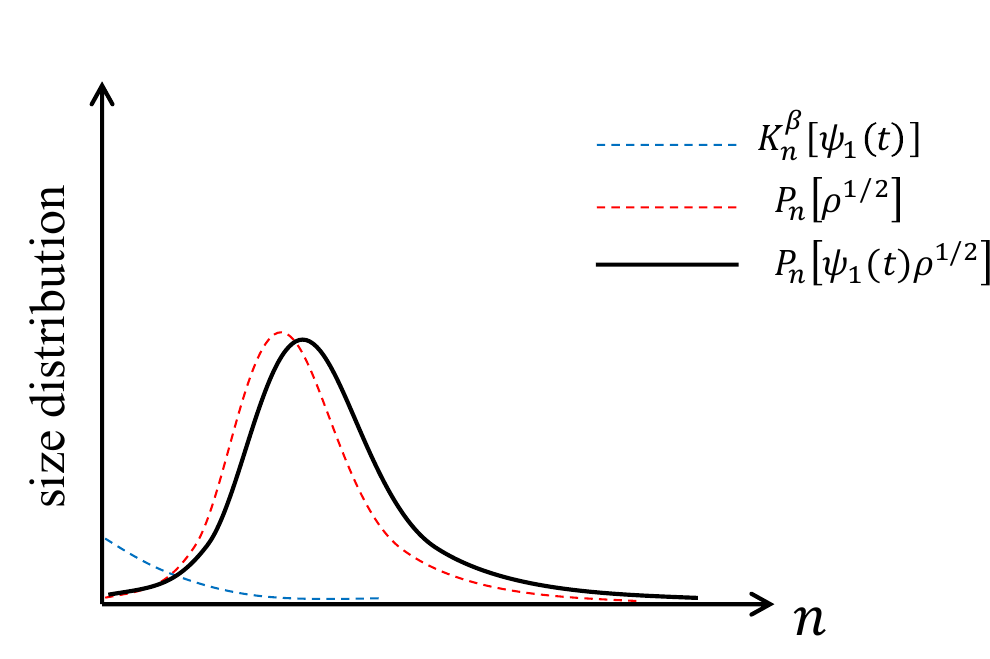} 
\par\end{centering}
\caption{Schematic illustration of the size distribution $P_{n}\left[\psi_{1}\left(t\right)\rho^{1/2}\right]$
for the operator $\psi_{1}\left(t\right)\rho^{1/2}$ (black curve)
which naturally decomposes into a convolution of a growth distribution
$K_{n}^{\beta}\left[\psi_{1}\left(t\right)\right]$ (blue dashed curve)
with the size distribution $P_{n}\left[\rho^{1/2}\right]$ of the
operator $\rho^{1/2}$ (red dashed curve). This is due to the factorization
relation of their respective generating functions (\ref{eq:Near-Gibbs Generator}).}
\end{figure}
Note that the discussion above can be generalized to arbitrary operators.
For an arbitrary operator $\mathcal{O}$, as long as we normalize
it such that $\langle TFD|\mathcal{O}_{L}^{\dagger}\mathcal{O}_{L}|TFD\rangle\equiv\left\langle \mathcal{O}^{\dagger}\mathcal{O}\right\rangle _{\beta}=1$,
the expansion of the two-point function 
\begin{equation}
\mathcal{G}_{\mu}\left[\mathcal{O}\right]\equiv\frac{\mathcal{Z}_{\mu}\left[\mathcal{O}\rho^{1/2}\right]}{\mathcal{Z}_{\mu}\left[\rho^{1/2}\right]}\label{eq:generaltwopointfunc}
\end{equation}
measures the effective size distribution of $\mathcal{O}$ when applied
to the thermal state.

\subsection{Twisted Boundary Condition}

We are interested in studying the generating function $\mathcal{Z}_{\mu}\left[\mathcal{O}\right]$
for $\mathcal{O}=\rho^{1/2}$ and $\mathcal{O}=\psi_{1}(t)\rho^{1/2}$.
As we discussed earlier, inserting the operator $\exp\left(-\mu n\right)$
corresponds to changing the imaginary time evolution. The computation
can be simplified by noticing that $\exp\left(-\mu n\right)$ is a
Gaussian operator, such that its action by conjugation to fermion
operators $\psi_{i}^{L,R}$ leads to a simple linear superposition:
\begin{equation}
e^{\mu n}\left(\begin{array}{c}
\psi^{L}\\
i\psi^{R}
\end{array}\right)e^{-\mu n}=\left(\begin{array}{cc}
\cosh\left(\mu\right) & -\sinh\left(\mu\right)\\
-\sinh\left(\mu\right) & \cosh\left(\mu\right)
\end{array}\right)\left(\begin{array}{c}
\psi^{L}\\
i\psi^{R}
\end{array}\right)\label{eq:Op Bdry Cond}
\end{equation}
As a result, inserting the operator-weighting term $\exp\left(-\mu n\right)$
is equivalent to twisting the boundary condition of the fermion fields
at $\tau=\beta/4$.

It is convenient to ``de-purify'' the system and return to the single
copy of fermion fields, but with a twisted boundary condition. The
single field is defined by continuously stitching the left and right
fields together: 
\begin{equation}
\psi_{i}\left(\tau\right)=\begin{cases}
\psi_{i}^{L}\left(\tau\right) & 0\leq\tau\leq\beta/2\\
i\psi_{i}^{R}\left(\beta-\tau\right) & \beta/2\leq\tau<\beta
\end{cases}\label{eq:De-Purify}
\end{equation}
with the requirement of course that $\psi\left(\tau+\beta\right)=-\psi\left(\tau\right)$.
This stitching transforms the purified action for the two fields into
the original action for this single field; however, the twist condition
must accompany the fields. In conclusion, the two-sided path integral
in the presence of the factor $\exp\left(-\mu n\left(\beta/4\right)\right)$
equals the original path integral where the fields are twisted according
to 
\begin{align}
\lim_{\tau\rightarrow\beta/4^{+}}\left(\begin{array}{c}
\psi\left(\tau\right)\\
\psi\left(\beta-\tau\right)
\end{array}\right) & =\left(\begin{array}{cc}
\cosh\left(\mu\right) & -\sinh\left(\mu\right)\\
-\sinh\left(\mu\right) & \cosh\left(\mu\right)
\end{array}\right)\lim_{\tau\rightarrow\beta/4^{-}}\left(\begin{array}{c}
\psi\left(\tau\right)\\
\psi\left(\beta-\tau\right)
\end{array}\right)\label{eq:Bdry Conds}
\end{align}
Therefore, we conclude that calculating the two point function $\mathcal{G}_{\mu}$
is equivalent to calculating the original two-point function, but
with the following twisted boundary conditions 
\begin{equation}
\left(\begin{array}{c}
\underset{\tau_{1/2}\rightarrow\beta/4^{+}}{\lim}\mathcal{G}_{\mu}\left(\tau_{1},\tau_{2}\right)\\
\underset{\tau_{1/2}\rightarrow3\beta/4^{-}}{\lim}\mathcal{G}_{\mu}\left(\tau_{1},\tau_{2}\right)
\end{array}\right)=\left(\begin{array}{cc}
\cosh\left(\mu\right) & -\sinh\left(\mu\right)\\
-\sinh\left(\mu\right) & \cosh\left(\mu\right)
\end{array}\right)\left(\begin{array}{c}
\underset{\tau_{1/2}\rightarrow\beta/4^{-}}{\lim}\mathcal{G}_{\mu}\left(\tau_{1},\tau_{2}\right)\\
\underset{\tau_{1/2}\rightarrow3\beta/4^{+}}{\lim}\mathcal{G}_{\mu}\left(\tau_{1},\tau_{2}\right)
\end{array}\right)\label{eq:Two-Point Bdry Conds}
\end{equation}

\begin{figure}[tb]
\centering{}\includegraphics[width=1\textwidth]{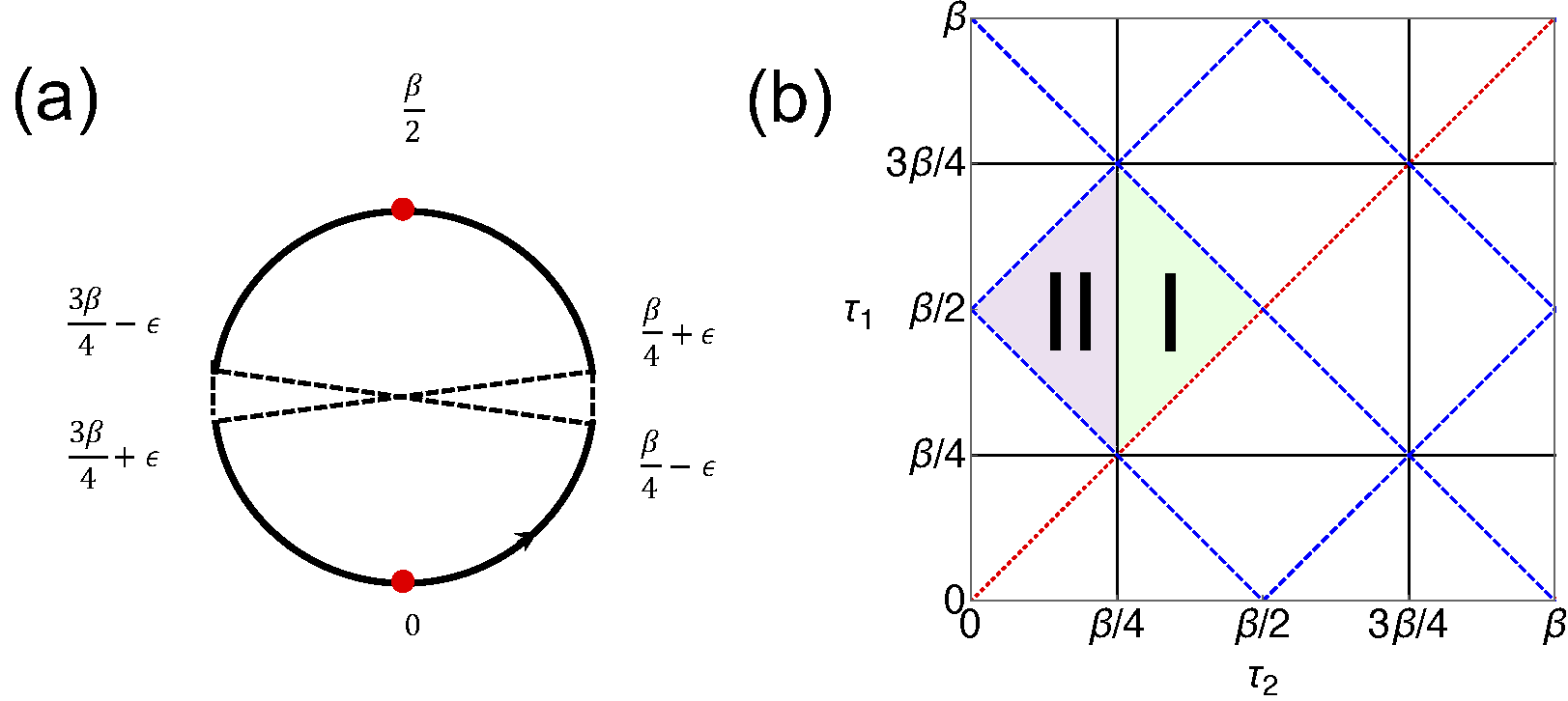}
\caption{(a) The twisted boundary condition on the imaginary time circle. When
$\tau$ crosses $\beta/4$ from below, $\psi(\tau)$ becomes a superposition
of $\psi\left(\beta/4+\epsilon\right)$ and $\psi\left(3\beta/4-\epsilon\right)$
(see Eq.\ (\ref{eq:Bdry Conds}). (b) The various symmetry and boundary
conditions on the twisted two point function in the $\left(\tau_{1},\tau_{2}\right)$
plane. First, $\mathcal{G}_{\mu}$ is odd under reflections across
the red dotted line and even under reflections across the blue dashed
lines. Thus, it is sufficient to solve the saddle-point equations
in the fundamental domain $0<\tau_{1}-\tau_{2}<\beta/2$ and $\beta/2<\tau_{1}+\tau_{2}<\beta$.
The black lines are the locations of the twisting boundary conditions
(\ref{eq:Two-Point Bdry Conds}), which reduce in the large $q$ limit
(\ref{eq:Large q Bdry Con}) and divide our fundamental domain into
two regions. Region I is where neither of the two fermions have crossed
a twist, while in Region II the fermions are on opposite sides of
the twist. \label{fig:boundary condition}}
\end{figure}
We note that while these conditions break time translation invariance,
they preserve a set of discrete symmetries. Specifically, if the original
Hamiltonian is time-reversal invariant, then $\mathcal{G}_{\mu}\left(\tau_{1},\tau_{2}\right)$
has reflection symmetry across the lines $\tau_{1}\pm\tau_{2}=n\beta/2$
for all integers $n\in\mathbb{Z}$ 
\begin{equation}
\mathcal{G}_{\mu}\left(\tau_{1},\tau_{2}\right)=\mathcal{G}_{\mu}\left(\frac{n\beta}{2}-\tau_{2},\frac{n\beta}{2}-\tau_{1}\right)=\left(-1\right)^{n+1}\mathcal{G}_{\mu}\left(\tau_{2}+\frac{n\beta}{2},\tau_{1}-\frac{n\beta}{2}\right)\label{eq:Reflection Conditions}
\end{equation}
Thus, we need only to solve for $\mathcal{G}_{\mu}\left(\tau_{1},\tau_{2}\right)$
in the fundamental domain $0<\tau_{1}-\tau_{2}<\beta/2$ and $\beta/2<\tau_{1}+\tau_{2}<\beta$,
as shown in Fig. \ref{fig:boundary condition}(b) by the union of
regions I and II.

\subsection{Thermally Renormalized Unit of Size\label{subsec:Thermal-Renormalization}}

As an interesting application of our formalism, let us note how $\psi_{1}(t)$
affects $\rho^{1/2}$ by taking $t=0$ and consider the change of
average size by a single fermion operator $\psi_{1}$. 
\begin{align}
\Delta n_{\beta}\left[\psi_{1}\right] & \equiv n\left[\psi_{1}\rho^{1/2}\right]-n\left[\rho^{1/2}\right]\nonumber \\
 & =\frac{1}{2}\sum_{i=1}^{N}\left(\left\langle TFD\right|\psi_{1}^{L}i\psi_{i}^{L}\psi_{i}^{R}\psi_{1}^{L}\left|TFD\right\rangle -\left\langle TFD\right|i\psi_{i}^{L}\psi_{i}^{R}\left|TFD\right\rangle \right)\nonumber \\
 & =\left\langle TFD\right|i\psi_{1}^{R}\psi_{1}^{L}\left|TFD\right\rangle =G_{11}\left(\frac{\beta}{2}\right)\label{eq:renormalized size}
\end{align}
At infinite temperature $\beta\rightarrow0$, $G\left(\beta/2\right)=1$,
which restores the trivial result that $\psi_{1}$ increases the size
of the density operator (which is proportional to identity operator,
with size $0$) by $1$. At finite temperature, interestingly, the
size change induced by a single fermion operator is smaller than $1$,
and is given by the same imaginary time two-point function as the
one that determines the fractional scrambling distance $\delta_{\beta}=1-\frac{n\left[\rho^{1/2}\right]}{N/2}$
in Eq. (\ref{eq:Average Thermal State Distance Two-Point Function}).
In general, the size increase induced by $\psi_{i}$ is $\Delta n_{\beta}\left[\psi_{i}\right]=G_{ii}\left(\beta/2\right)$,
which may depend on $i$. The average size increase is exactly $\delta_{\beta}$.
\footnote{In term of the probabilities $K_{m}^{\beta}\left[\psi_{1}\right]$
in Eq. (\ref{eq:moments of G}), we have $\Delta n\left[\psi_{i}\right]=\sum_{m=0}^{N}mK_{m}^{\beta}\left[\psi_{1}\right]$. }
\begin{equation}
\frac{1}{N}\sum_{i}\Delta n\left[\psi_{i}\right]=\frac{1}{N}\sum_{i}G_{ii}\left(\frac{\beta}{2}\right)=G\left(\frac{\beta}{2}\right)=\delta_{\beta}\label{eq:average size increase}
\end{equation}

\begin{figure}[tb]
\begin{centering}
\includegraphics[width=0.22\textheight]{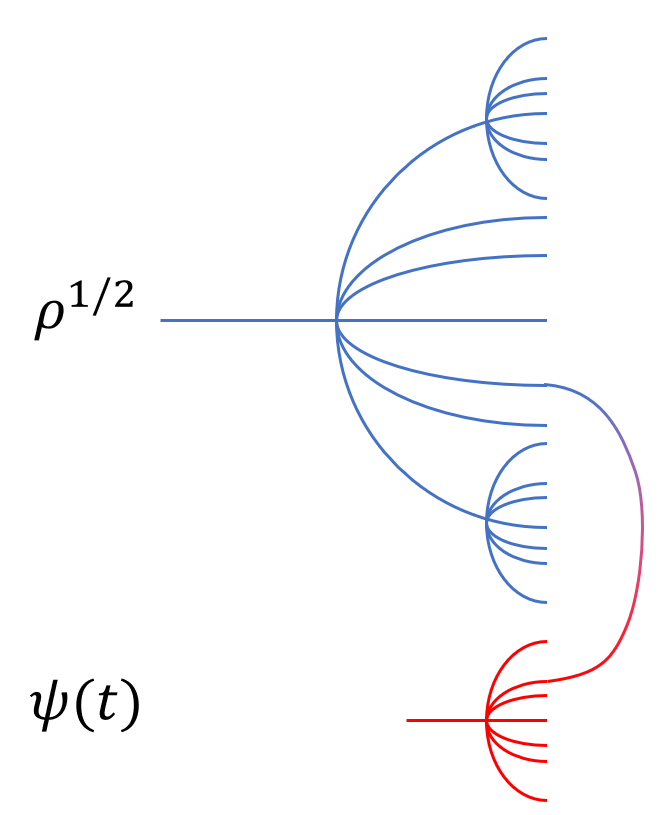} 
\par\end{centering}
\caption{\label{fig:collision}At finite temperature, when an operator such
as $\psi_{1}\left(t\right)$ is multiplied to $\rho^{1/2}$, there
is a chance that some fermion flavors collide and the size increase
is smaller than the size of $\psi_{1}\left(t\right)$ itself.}
\end{figure}

Physically, the average size change due to applying a single fermion
is generically $\delta_{\beta}<1$ at finite $\beta$, because in
the presence of a nontrivial $\rho^{1/2}$ there is a chance that
multiplying by $\psi_{1}$ decreases the size, as is illustrated in
Fig. \ref{fig:collision}, although the chance of increasing the size
is always bigger. The closer the length of $\rho^{1/2}$ is to the
scrambling value $n_{*}=N/2$, the smaller is the size increase $\Delta n_{\beta}\left[\psi_{1}\rho^{1/2}\right]$.
For a fully scrambled operator with $n=N/2,~\delta=0$, multiplying
a fermion $\psi_{1}$ has equal chance of increasing or decreasing
the size, so that the average size stays the same.

It should be emphasized that the discussion above is not restricted
to the thermal density operator. For any operator $\mathcal{O}$,
we can define the size change 
\[
\Delta n_{\mathcal{O}}\left[\psi_{i}\right]\equiv n\left[\psi_{i}\mathcal{O}\right]-n\left[\mathcal{O}\right]
\]
and obtain the following identity: 
\[
\frac{1}{N}\sum_{i}\Delta n_{\mathcal{O}}\left[\psi_{i}\right]=\delta\left[\mathcal{O}\right]\equiv1-\frac{2n\left[\mathcal{O}\right]}{N}
\]
The only thing special for the thermal density operator is the relation
of $\Delta n$ to imaginary time two-point function in a single-copy
system.

Furthermore, instead of $\psi_{i}$ we can consider a string $\Gamma_{I}\equiv\Gamma_{i_{1}i_{2}...i_{k}}=i^{\frac{k\left(k-1\right)}{2}}\psi_{i_{1}}...\psi_{i_{k}}\qquad1\leq i_{1}<i_{2}<...<i_{k}\leq N$
introduced in Eq. (\ref{eq:majoranastring}), and consider how the
size of $\Gamma_{I}\mathcal{O}$ is different from $\mathcal{O}$.
We have 
\begin{align}
\Delta n_{\mathcal{O}}\left[\Gamma_{I}\right] & \equiv n\left[\Gamma_{I}\mathcal{O}\right]-n\left[\mathcal{O}\right]=\sum_{s=1}^{k}\left\langle \mathcal{O}\right|i\psi_{i_{s}}^{R}\psi_{i_{s}}^{L}\left|\mathcal{O}\right\rangle \label{eq:size change general operator}
\end{align}
If we average over all Majorana strings $\Gamma_{I}$ with the same
size $k$, we obtain 
\begin{align}
\frac{1}{C_{N}^{k}}\sum_{I}\Delta n_{\mathcal{O}}\left[\Gamma_{I}\right] & =k\delta\left[\mathcal{O}\right]
\end{align}
In the last equation, $C_{N}^{k}=\frac{N!}{k!(N-k)!}$ is the number
of strings with length $k$. This equation shows that the average
size change induced by multiplying a string with length $k$ is $k$
times $\delta\left[\mathcal{O}\right]$, further confirms that each
fermion in the string contributes additively.

This observation suggests that at finite temperature (or more generally,
for any density operator $\rho$), the fractional scrambling distance
$\delta$, rather than $1$, should be considered as the fundamental
unit of size, which is carried by each fermion operator. Indeed, as
we will discuss in next section, our calculation in the SYK model
in the large $q$ limit suggests universal behavior occurs when size
is measured in this renormalized unit.

\section{SYK Model\label{sec:SYK-Model}}

In this section, we will study the operator size growth in the SYK
model \cite{Sachdev:1992fk,Kitaev:2014t2}. This model features $q$-local
interactions with independently random couplings, where each of the
couplings is normal distributed 
\[
H=i^{q/2}\sum_{1\leq i_{1}...\leq i_{q}\leq N}J_{i_{1}...i_{q}}\psi_{i_{1}}...\psi_{i_{q}}\qquad\left\langle J_{i_{1}...i_{q}}^{2}\right\rangle =\frac{J^{2}}{\binom{N-1}{q-1}}=\frac{\mathcal{J}^{2}}{2q\binom{N-1}{q-1}}\qquad\left\{ \psi_{i},\psi_{j}\right\} =2\delta_{ij}
\]
At large $N$, the two-point function satisfies the saddle-point equations
\begin{align}
\left[G\right]^{-1} & =[G_{0}]^{-1}-\left[\Sigma\right]\qquad\Sigma\left(\tau_{1},\tau_{2}\right)=\frac{\mathcal{J}^{2}}{2q}\left(G\left(\tau_{1},\tau_{2}\right)\right)^{q-1}\label{eq:Orig Sad-Pt}
\end{align}
where bracketed terms are Matsubara frequency matrices. One should
note that since the fermions square to one, $\left[G_{0}\right]^{-1}=-i\omega/2$
rather than $-i\omega$.

\subsection{Large $q$ Approximation}

\begin{figure}[tb]
\begin{centering}
\includegraphics[width=1\textwidth]{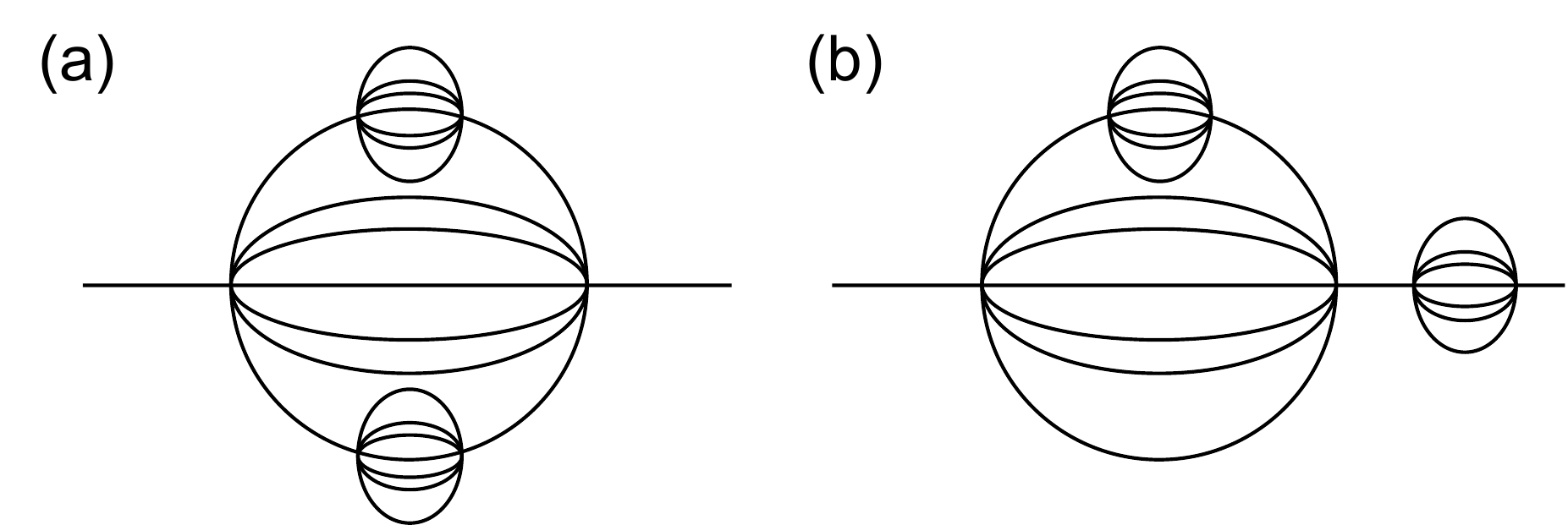} 
\par\end{centering}
\caption{Two $\mathcal{O}\left(J^{6}\right)$ examples of the planar graphs
of that survive the large-$N$ limit. Only graphs of form (a), where
melons are inserted into melons, survive the large $q$ limit. Notice
that there are $\mathcal{O}\left(q^{2}\right)$ graphs of form (a),
while there are only $\mathcal{O}\left(q\right)$ graphs of form (b).
This is because there are $\mathcal{O}\left(q\right)$ locations at
any depth of a given graph to insert another melon; however, there
are typically only $\mathcal{O}\left(1\right)$ locations to thread
another melon. Accordingly, we take our coupling $J$ to equal $\mathcal{J}/\sqrt{2q}$.
Consequently, the $\mathcal{O}\left(q\right)$ combinatorial enhancement
gained for each new melon insertion is canceled by the $J^{2}=\mathcal{J}^{2}/\left(2q\right)$
factor accompanying said melon. This $q$-scaling of the coupling
isolates the infinite subset of the planar graphs where the graphs
are two copies of a tree that are then glued together (a.k.a. ``doubletree''
graphs) such as (a). All non-doubletree graphs such as (b) are suppressed
in $q$ since they receive factors of $\mathcal{J}^{2}/\left(2q\right)$
for each melon, but do not receive the necessary number of $q$ combinatorial
enhancements. \label{fig:large q diagrams}}
\end{figure}
In the language of Feynman diagrams, the Schwinger-Dyson equation
(\ref{eq:Orig Sad-Pt}) corresponds to only keeping the leading ``melon
diagrams'' as is shown in Fig. \ref{fig:large q diagrams}. All other
diagrams are sub-leading in large $N$. In the large $q$ limit, there
are two types of diagrams. Those with melons inserted into melons
(such as Fig. \ref{fig:large q diagrams}(a)) receive a combinatorial
$q$ enhancement, as there are many rungs upon which one may insert
(hence the need for a $q^{-1}$ factor in the self-energy to keep
everything finite). In contrast, diagrams where melons are simply
threaded together (such as Fig. \ref{fig:large q diagrams}(b)) do
not receive this enhancement \cite{Gross:2016kjj}. Thus, at large
$q$ only the former dominate, which corresponds to the following
truncation of the Schwinger-Dyson expansion: 
\begin{align*}
\left[G\right] & =\left[G_{0}\right]+\left[G_{0}\right]\left[\Sigma\right]\left[G_{0}\right]\qquad\Sigma\left(\tau_{1},\tau_{2}\right)=\frac{\mathcal{J}^{2}}{2q}\left(G\left(\tau_{1},\tau_{2}\right)\right)^{q-1}
\end{align*}

Combing the equations together and Fourier transforming, one obtains
\[
\partial_{\tau_{1}}\partial_{\tau_{2}}\left(G-G_{0}\right)=-\frac{2\mathcal{J}^{2}}{q}G^{q-1}
\]
The role played by $G_{0}$ in this equation is to require that $G\rightarrow G_{0}$
as $\tau_{12}$ goes to integer multiples of $\beta$. Therefore,
if we take $G=G_{0}e^{\sigma/q}$ with $\sigma\rightarrow0$ at the
$\tau_{12}$ boundaries, we obtain Liouville's equation\cite{Maldacena:2016hyu,eberlein2017quantum}
\begin{align}
\partial_{\tau_{1}}\partial_{\tau_{2}}\sigma & =-2\mathcal{J}^{2}e^{\sigma}+\mathcal{O}\left(1/q\right)\label{eq:Liouville}
\end{align}
where the field $\sigma$ is expected to be periodic in both of its
arguments, as well as have kinks when $\tau_{12}$ approaches integer
multiples of $\beta$.

Now in order to find $\mathcal{G}_{\mu}$, we will need to solve the
above equations with the twisted boundary conditions (\ref{eq:Two-Point Bdry Conds}).
Furthermore, our twisted two-point function $\mathcal{G}_{\mu}$ also
satisfies the reflection conditions (\ref{eq:Reflection Conditions}).
Thus, we need only solve for $\mathcal{G}_{\mu}$ in the fundamental
domain $0<\tau_{1}-\tau_{2}<\beta/2$ and $\beta/2<\tau_{1}+\tau_{2}<\beta$,
as shown in Fig. \ref{fig:boundary condition}(b) by the union of
regions I and II.

\subsection{The Large-$q$ solution}

In the large $q$ limit, each commutator with the Hamiltonian increases
the size of operator by $\sim q$, so that it is natural to measure
the operator size in unit of $q$. In the generating function, this
corresponds to defining $\hat{\mu}\equiv q\mu$ (for reasons to be
explained in the next subsection we will actually use the slightly
smaller variable $\hat{\mu}=q\delta_{\beta}\mu$ (\ref{eq:mu hat renormalization})),
with $\hat{\mu}$ kept finite in the large $q$ limit. The derivative
of the generating function over $\hat{\mu}$ measures the size $n$
in unit of $q$. If we consider the large $q$ limit with $\hat{\mu}$
being kept finite, and use the large-$q$ ansatz for twisted two-point
function 
\[
\mathcal{G}_{\mu}\left(\tau_{1},\tau_{2}\right)=G_{0}\left(\tau_{1},\tau_{2}\right)e^{\sigma_{\mu}\left(\tau_{1},\tau_{2}\right)/q},
\]
the boundary condition (\ref{eq:Two-Point Bdry Conds}) reduces to
\begin{align}
\left(\begin{array}{c}
\underset{\tau_{1/2}\rightarrow\beta/4^{+}}{\lim}\mathcal{G}_{\mu}\left(\tau_{1},\tau_{2}\right)\\
\underset{\tau_{1/2}\rightarrow3\beta/4^{-}}{\lim}\mathcal{G}_{\mu}\left(\tau_{1},\tau_{2}\right)
\end{array}\right) & \simeq\left(\begin{array}{cc}
1 & -\frac{\hat{\mu}}{q}\\
-\frac{\hat{\mu}}{q} & 1
\end{array}\right)\left(\begin{array}{c}
\underset{\tau_{1/2}\rightarrow\beta/4^{-}}{\lim}\mathcal{G}_{\mu}\left(\tau_{1},\tau_{2}\right)\\
\underset{\tau_{1/2}\rightarrow3\beta/4^{+}}{\lim}\mathcal{G}_{\mu}\left(\tau_{1},\tau_{2}\right)
\end{array}\right)\nonumber \\
 & \simeq e^{-\hat{\mu}/q}\left(\begin{array}{c}
\underset{\tau_{1/2}\rightarrow\beta/4^{-}}{\lim}\mathcal{G}_{\mu}\left(\tau_{1},\tau_{2}\right)\\
\underset{\tau_{1/2}\rightarrow3\beta/4^{+}}{\lim}\mathcal{G}_{\mu}\left(\tau_{1},\tau_{2}\right)
\end{array}\right)\label{eq:Large q Bdry Con}
\end{align}
Thus, to the leading order of $\frac{1}{q}$, the two equations for
$\beta/4$ and $3\beta/4$ decouple.

As a reminder, these equations encode the effect of moving a Majorana
fermion on the Euclidean circle (i.e. the two TFD half-circles) from
one side of the twist operator (\ref{eq:twopointfunc}) to the other.
The factor of $\exp\left(-\hat{\mu}/q\right)=\exp\left(-\mu\right)$
denotes the fact that such an action changes the size of the doubled
state by exactly one whole fermion. However, we shall find that for
states of large size such as the thermofield double (i.e. Gibbs state)
each Majorana fermion actually increases the total average operator
size by a smaller fraction $\delta_{\beta}$ rather than $1$, where
the size change $\delta_{\beta}$ is smaller when the state's size
is larger. As such, we shall find that it will be appropriate to use
the variable $\hat{\mu}=q\mu\delta_{\beta}$ instead of $\hat{\mu}=q\mu$
when taking the large-$q$ limit.

The twisted two-point function in large-$q$ limit can thus be obtained
by solving Liouville's equation (\ref{eq:Liouville}) with the $\mu$-dependent
boundary conditions (\ref{eq:Large q Bdry Con}). Here we will skip
the tedious details and directly present the solution. When the times
are such that the two fermions are on the same side of the twisted
boundary, which corresponds to $\tau_{2}>\beta/4$ (region I in Fig.
\ref{fig:boundary condition}(b), we find a seemingly time-translation
invariant solution solution 
\begin{align}
\mathcal{G}_{\mu}\left(\tau_{1},\tau_{2}\right) & =\left(\frac{\sin\gamma_{\mu}}{\sin\left(\alpha_{\mu}\left(\tau_{1}-\tau_{2}\right)+\gamma_{\mu}\right)}\right)^{2/q}\equiv G_{\mu}\left(\tau_{1}-\tau_{2}\right)\label{eq:twisted2ptfunc1}
\end{align}
However, when the times are such that the two fermions are on opposite
sides of the twisted boundary at $\beta/4$ and $3\beta/4$, which
for our domain amounts to the condition $\tau_{2}<\beta/4$ (region
II in Fig. \ref{fig:boundary condition}(b), the time translation
symmetry is explicitly broken 
\begin{align}
\mathcal{G}_{\mu}\left(\tau_{1},\tau_{2}\right) & =\frac{e^{-\hat{\mu}/q}G_{\mu}\left(\tau_{1}-\tau_{2}\right)}{\left(1-\frac{\left(1-e^{-\hat{\mu}}\right)}{\sin^{2}\gamma_{\mu}}\left(G_{\mu}\left(\tau_{1}-\tau_{2}\right)\right)^{q/2}\sin\left(\alpha_{\mu}\left(\tau_{1}-\frac{\beta}{4}\right)\right)\sin\left(\alpha_{\mu}\left(\tau_{2}-\frac{\beta}{4}\right)\right)\right)^{2/q}}\label{eq:twisted2ptfunc2}
\end{align}
Here the parameters $\alpha_{\mu}$ and $\gamma_{\mu}$ are functions
of $\beta\mathcal{J}$ and $\mu$, which are determined by the boundary
condition $\mathcal{G}(\tau,\tau)=1$ as well as the reflection conditions
(\ref{eq:Reflection Conditions}) 
\begin{align}
\alpha_{\mu}\beta & =\beta\mathcal{J}\sin\gamma_{\mu},\quad\sin\left(\frac{\alpha_{\mu}\beta}{2}+2\gamma_{\mu}\right)=e^{-\hat{\mu}}\sin\left(\frac{\alpha_{\mu}\beta}{2}\right)\label{eq:Full Parameter Constraints}
\end{align}
In the limit $\mu\rightarrow0$, we recover the untwisted two-point
function $\mathcal{G}_{\mu=0}(\tau_{1},\tau_{2})=G_{\mu=0}(\tau_{1}-\tau_{2})=G\left(\tau_{1}-\tau_{2}\right)$
in the whole domain, and the equation for the parameters reduce to
the ordinary case\cite{Maldacena:2016hyu}: 
\begin{equation}
\alpha_{\mu=0}\equiv\alpha=\mathcal{J}\cos\left(\frac{\alpha\beta}{2}\right),\quad\gamma_{\mu=0}\equiv\gamma=\frac{\pi-\alpha\beta}{2}\label{eq:Parameter Constraints}
\end{equation}
The asymptotic behavior at small values of $\beta\mathcal{J}$ and
large values of $\beta\mathcal{J}$ respectively are given by 
\begin{align*}
\alpha & =\mathcal{J}\left(1-\frac{\beta^{2}\mathcal{J}^{2}}{8}+\mathcal{O}\left(\beta^{4}\mathcal{J}^{4}\right)\right) & \alpha & =\frac{\pi}{\beta}\left(1-\frac{2}{\beta\mathcal{J}}+\mathcal{O}\left(\frac{1}{\beta^{2}\mathcal{J}^{2}}\right)\right)
\end{align*}

\subsection{Size renormalization}

Before carrying further analysis to the SYK operator growth in next
section, we need to discuss an important modification to the two-point
function solution due to higher order $q$ effects. If we take $\tau_{1}\rightarrow\beta/4+\epsilon,~\tau_{2}\rightarrow\beta/4-\epsilon$
in Eq. (\ref{eq:twisted2ptfunc2}), we obtain
\begin{equation}
\mathcal{G}_{\mu}\left(\frac{\beta}{4}+\epsilon,\frac{\beta}{4}-\epsilon\right)=e^{-\hat{\mu}/q}\label{eq:twopt zero time}
\end{equation}
This is the kernel that determines the size change induced by multiplying
$\psi_{1}$ to $\rho^{1/2}$, which has been discussed in section
(\ref{subsec:Thermal-Renormalization}). Taking the $\mu$-derivative
of $\mathcal{G}_{\mu}$, we find
\begin{align*}
\Delta n_{\beta}\left[\psi_{1}\right] & \equiv n\left[\psi_{1}\rho^{1/2}\right]-n\left[\rho^{1/2}\right]=-\left.\partial_{\mu}\log\mathcal{G}_{\mu}\left(\frac{\beta}{4}+\epsilon,\frac{\beta}{4}-\epsilon\right)\right|_{\mu=0}=1
\end{align*}
However, we also know that the size change is directly determined
by the two-point function due to Eq. (\ref{eq:renormalized size}):
\[
\Delta n_{\beta}\left[\psi_{1}\right]=\delta_{\beta}=G\left(\frac{\beta}{2}\right)=\left(\frac{\alpha}{\mathcal{J}}\right)^{2/q}
\]
where $\alpha\equiv\alpha_{\mu=0}\left(\beta\mathcal{J}\right)$ is
the smallest positive root of Eq. (\ref{eq:Parameter Constraints}). 

This discrepancy between the two calculations is because $\delta_{\beta}\rightarrow1$
in the large $q$ limit, and the $\mathcal{O}\left(q^{-1}\right)$
difference is neglected in the approximation we made to the boundary
condition. The easiest way to resolve this issue and makes a consistent
large-$q$ limit is by redefining $\hat{\mu}=q\mu$ to
\begin{equation}
\hat{\mu}=q\mu\delta_{\beta}\label{eq:mu hat renormalization}
\end{equation}
in Eq. (\ref{eq:Large q Bdry Con}), which leads to the same substitution
in Eqs. (\ref{eq:twisted2ptfunc1}), (\ref{eq:twisted2ptfunc2}),
and (\ref{eq:Full Parameter Constraints}). In the following, we will
always use this definition of $\hat{\mu}$.

Physically, this substitution is a consequence of the size renormalization
discussed in section (\ref{subsec:Thermal-Renormalization}). Each
Majorana fermion increase the operator size by $\delta_{\beta}$ rather
than $1$. Each action of the Hamiltonian increases the operator size
by $\sim q\delta_{\beta}$. Although in large $q$ limit $1-\delta_{\beta}$
is order $q^{-1}$, it is important to keep track of this distance,
since the same $\delta_{\beta}$ also measures the fractional scrambling
distance of $\rho^{1/2}$, as we discussed in section (\ref{subsec:thermal state}).
The size of $\rho^{1/2}$ is 
\[
n\left[\rho^{1/2}\right]=\frac{N}{2}\left(1-\delta_{\beta}\right)=\frac{N}{2}\left(1-\left(\frac{\alpha}{\mathcal{J}}\right)^{2/q}\right)
\]
which decreases with increasing $q$, but is always large since we
should always take the large $N$ limit before taking large $q$.

\section{SYK Operator Growth\label{sec:SYK-Operator-Growth}}

We are now equipped with everything we need to understand $P_{n}\left[\psi_{1}\left(t\right)\rho^{1/2}\right]$,
the size distribution of $\psi_{1}\left(t\right)\rho^{1/2}$. According
to Eq. (\ref{eq:Near-Gibbs Generator}), the generating function $\mathcal{Z}_{\mu}\left[\psi_{1}(t)\rho^{1/2}\right]$
for this distribution splits into a product of the thermal state's
generating function $\mathcal{Z}_{\mu}\left[\rho^{1/2}\right]$ and
$\mathcal{G}_{\mu}\left(\beta/4^{+}+it,\beta/4^{-}+it\right)$. The
latter is simply the twisted two-point function we discussed in the
previous subsection with an analytic continuation.

\subsection{Thermal State \label{subsec:SYK Thermal-State}}

The generating function $\mathcal{Z}_{\mu}\left[\rho^{1/2}\right]$
is the partition function of the system with the insertion $\exp\left(-\mu n\left(\beta/4\right)\right)$
divided by that of the original system (see Eq. (\ref{eq:thermalZmuTFD})).
This quantity can be determined by the twisted two-point function,
since one has 
\begin{align*}
-\partial_{\mu}\ln\mathcal{Z}_{\mu} & =\frac{\ev*{ne^{-\mu n}}{\rho^{1/2}}}{\ev*{e^{-\mu n}}{\rho^{1/2}}}=\frac{N}{2}\left(1-G_{\mu}\left(\frac{\beta}{2}\right)\right)=\frac{N}{2}\left(1-\frac{\sin^{2/q}\gamma_{\mu}}{\sin^{2/q}\left(\alpha_{\mu}\frac{\beta}{2}+\gamma_{\mu}\right)}\right)
\end{align*}
In theory, we can integrate this equation to obtain $\mathcal{Z}_{\mu}\left[\rho^{1/2}\right]$.
However, many important properties of the distribution can be inferred
from just the first and second moment.

\begin{figure}[tb]
\noindent \includegraphics[width=0.5\textwidth]{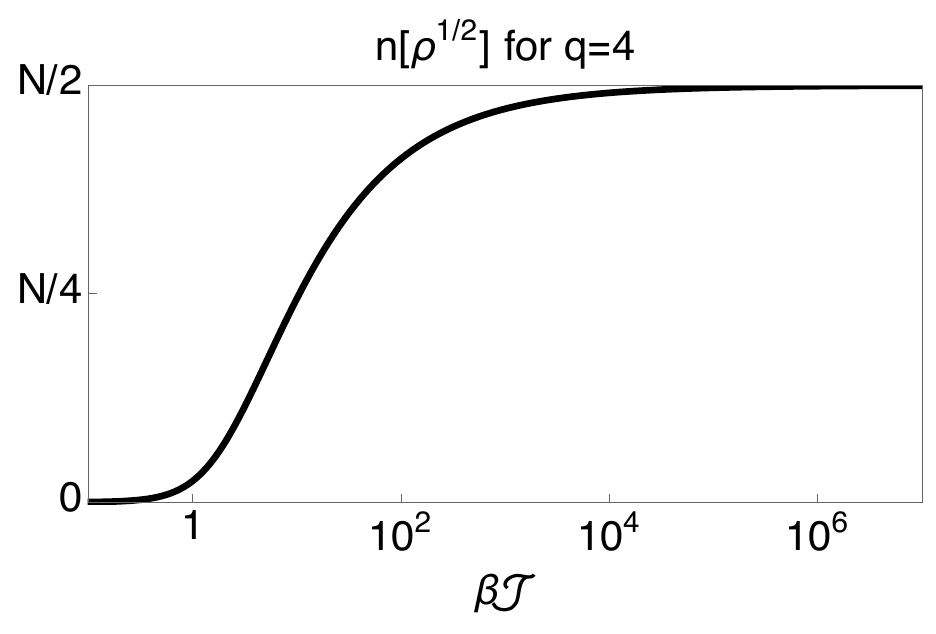}\includegraphics[width=0.5\textwidth]{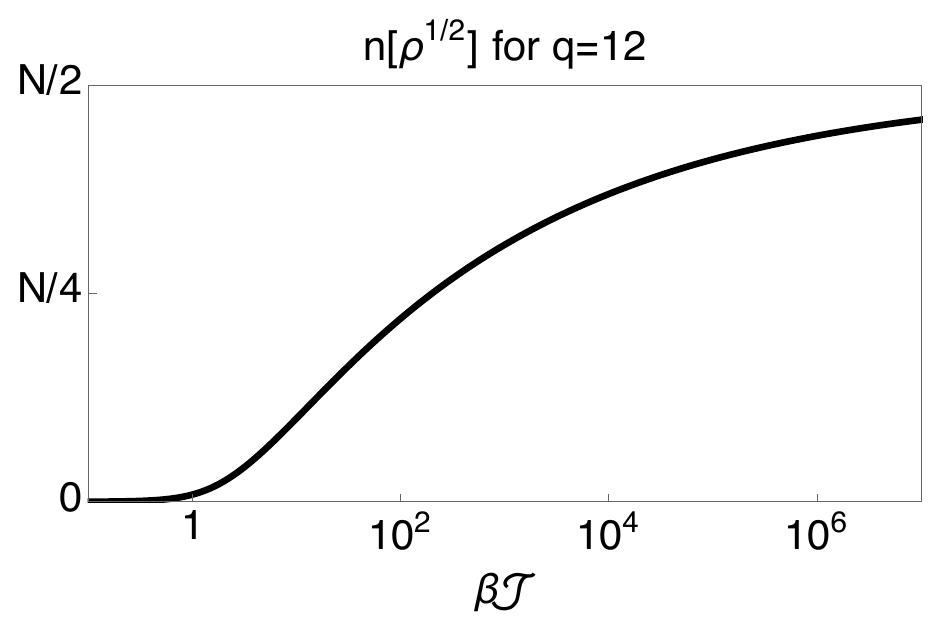}\caption{Plots of the average size of $\rho^{1/2}\propto\exp\left(-\beta H/2\right)$
for different $q$, given by Eq. (\ref{eq:Average Thermal Size}).
\label{fig:nbetaSYK}}
\end{figure}

The first moment is simply the average size 
\begin{equation}
n\left[\rho^{1/2}\right]=\frac{N}{2}\left(1-\delta_{\beta}\right)=\frac{N}{2}\left(1-G\left(\frac{\beta}{2}\right)\right)=\frac{N}{2}\left(1-\left(\frac{\alpha}{\mathcal{J}}\right)^{2/q}\right)\label{eq:Average Thermal Size}
\end{equation}
The behavior of $n\left[\rho^{1/2}\right]$ for two different $q$
values are plotted in Fig. \ref{fig:nbetaSYK} as a function of $\beta\mathcal{J}$.
Interestingly, we see that for larger $q$ it takes larger values
of $\beta\mathcal{J}$ to achieve the same average size. Thus, the
exponentiation of increasingly heavy SYK Hamiltonians results in relatively
\emph{lighter} thermal states. This is unintuitive, so one might argue
that it is simply due to the $q$-scaling nature of $\mathcal{J}$,
but that is only a constant shift in the log scale plot in Fig. \ref{fig:nbetaSYK},
which is nowhere large enough to account for the above discrepancy.
The true origin of this effect is the power of $2/q$ in the two-point
function. We conclude that this heavy-light relationship is thus a
non-trivial consequence of the large-$N$ and large-$q$ limit.

For $\beta\mathcal{J}\gg e^{q}$, one expects that the higher order
corrections to Liouville's equation (\ref{eq:Liouville}) cannot be
neglected, and so one should turn to the conformal approximation\cite{Maldacena:2016hyu}.
Their expression for the two-point function implies that the average
size of $\rho^{1/2}$ when $N\gg\beta\mathcal{J}\gg1$ is given by
\begin{align*}
n\left[\rho^{1/2}\right] & \approx\frac{N}{2}\left(1-c\left(q\right)\left(\frac{\pi}{\beta\mathcal{J}}\right)^{2/q}\right) & c\left(q\right) & =\left(\frac{\left(q-2\right)\tan\frac{\pi}{q}}{\pi}\right)^{1/q}
\end{align*}
The difference between the large $q$ and low temperature $n\left[\rho^{1/2}\right]=\frac{N}{2}\left(1-G\left(\frac{\beta}{2}\right)\right)$
is captured by the factor $c\left(q\right)$. It monotonically increases
from $c\left(4\right)=\left(2/\pi\right)^{1/4}\approx0.9$ when $q=4$,
and asymptotically approaches $1$ when $q$ is large as $1-2/q^{2}$.
We expect that $q=4$ and large $\beta\mathcal{J}$ will be where
our large $q$ approximation will have the largest error. However,
that this error is at worst $10\%$ renews our confidence that the
large $q$ approximation captures important analytic features of the
large-$N$ SYK model.

The second derivative of $\ln\mathcal{Z}_{\mu}$ determines the width
of the distribution: 
\begin{align*}
\sigma_{n}^{2}\left[\rho^{1/2}\right] & =\lim_{\mu\rightarrow0}\partial_{\mu}^{2}\ln\mathcal{Z}_{\mu}\left[\rho^{1/2}\right]=\frac{N}{2}\left.\partial_{\mu}G_{\mu}\left(\frac{\beta}{2}\right)\right|_{\mu\rightarrow0}\propto N
\end{align*}
Therefore the width of the distribution $\sigma_{n}\propto\sqrt{N}$,
such that the relative deviation from the average value $\sigma_{n}\left[\rho^{1/2}\right]/n\left[\rho^{1/2}\right]\propto N^{-1/2}$
is sharply peaked in the large $N$ limit. This is a consequence of
large $N$ factorization.

\subsection{Thermal Fermion}

As explicitly discussed in section (\ref{subsec:Thermal-Fermion}),
the generating function for the growth distribution $K^{\beta}$ (\ref{eq:moments of G})
is determined by the twisted two-point function (\ref{eq:twisted2ptfunc2})
\begin{align*}
\mathcal{G}_{\mu}\left(\frac{\beta^{+}}{4}+it,\frac{\beta^{-}}{4}+it\right) & =e^{-\mu\delta_{\beta}}\left(1+\left(1-e^{-q\mu\delta_{\beta}}\right)\left(\frac{\mathcal{J}}{\alpha_{\mu}}\sinh\alpha_{\mu}t\right)^{2}\right)^{-2/q}
\end{align*}
where $\alpha_{\mu}$ and $\gamma_{\mu}$ depend on $\mu$ and $\beta\mathcal{J}$
through the constraints (\ref{eq:Full Parameter Constraints}).

\subsubsection{Average Size}

This implies that the average size of the operator $\psi\left(t\right)\rho^{1/2}$
is given by 
\begin{align}
n\left[\psi_{1}\left(t\right)\rho^{1/2}\right] & =n\left[\rho^{1/2}\right]-\left.\partial_{\mu}\ln\mathcal{G}_{\mu}\left(\frac{\beta^{+}}{4}+it,\frac{\beta^{-}}{4}+it\right)\right|_{\mu=0}\nonumber \\
\Rightarrow n\left[\psi_{1}\left(t\right)\rho^{1/2}\right] & =\frac{N}{2}\left(1-\delta_{\beta}\right)+\delta_{\beta}\left(1+2\left(\frac{\mathcal{J}}{\alpha}\sinh\alpha t\right)^{2}\right)\label{eq:sizegrowthSYK}
\end{align}
where $\delta_{\beta}=\left(\alpha/\mathcal{J}\right)^{2/q}$, and
$\alpha\equiv\alpha_{\mu=0}\left(\beta\mathcal{J}\right)$ is the
smallest positive root of Eq. (\ref{eq:Parameter Constraints}). We
see that the difference in averages sizes of $\psi\left(t\right)\rho^{1/2}$
and $\rho^{1/2}$ is a simple when expressed in the renormalized size
unit $\delta_{\beta}$, which inspires us to define a notion of the
``average growth'' of $\psi_{1}\left(t\right)$ as 
\begin{equation}
\Delta\tilde{n}_{\beta}\left[\psi_{1}(t)\right]\equiv\frac{n\left[\psi_{1}(t)\rho^{1/2}\right]-n\left[\rho^{1/2}\right]}{\delta_{\beta}}=1+2\left(\frac{\mathcal{J}}{\alpha}\sinh\alpha t\right)^{2}\label{eq:Average Growth}
\end{equation}

Now, scrambling occurs when the average size of $\psi_{1}\left(t\right)\rho^{1/2}$
given by Eq. (\ref{eq:sizegrowthSYK}) reaches $n_{*}=N/2$. This
produces a slightly complicated expression for the scrambling time
$t_{*}$; however, it simplifies dramatically when phrased in terms
of the average growth of $\psi_{1}\left(t\right)$. Manipulating the
scrambling time equation $n\left[\psi_{1}\left(t_{*}\right)\rho^{1/2}\right]=N/2$,
we find that one may equivalently state that scrambling occurs when
the average growth of $\psi_{1}\left(t\right)$ reaches $n_{*}=N/2$
\begin{equation}
\Delta\tilde{n}_{\beta}\left[\psi_{1}(t_{*})\right]=1+2\left(\frac{\mathcal{J}}{\alpha}\sinh\alpha t_{*}\right)^{2}=\frac{N}{2}\label{eq:Average Growth Scrambles}
\end{equation}
This growth is consistent with the known result of large-$q$ Lyapunov
exponent \cite{Maldacena:2016hyu}. 
\[
\lambda_{L}=2\alpha
\]

\subsubsection{Full Growth Structure}

\begin{figure}[tb]
\noindent \begin{centering}
\includegraphics[width=0.8\textwidth]{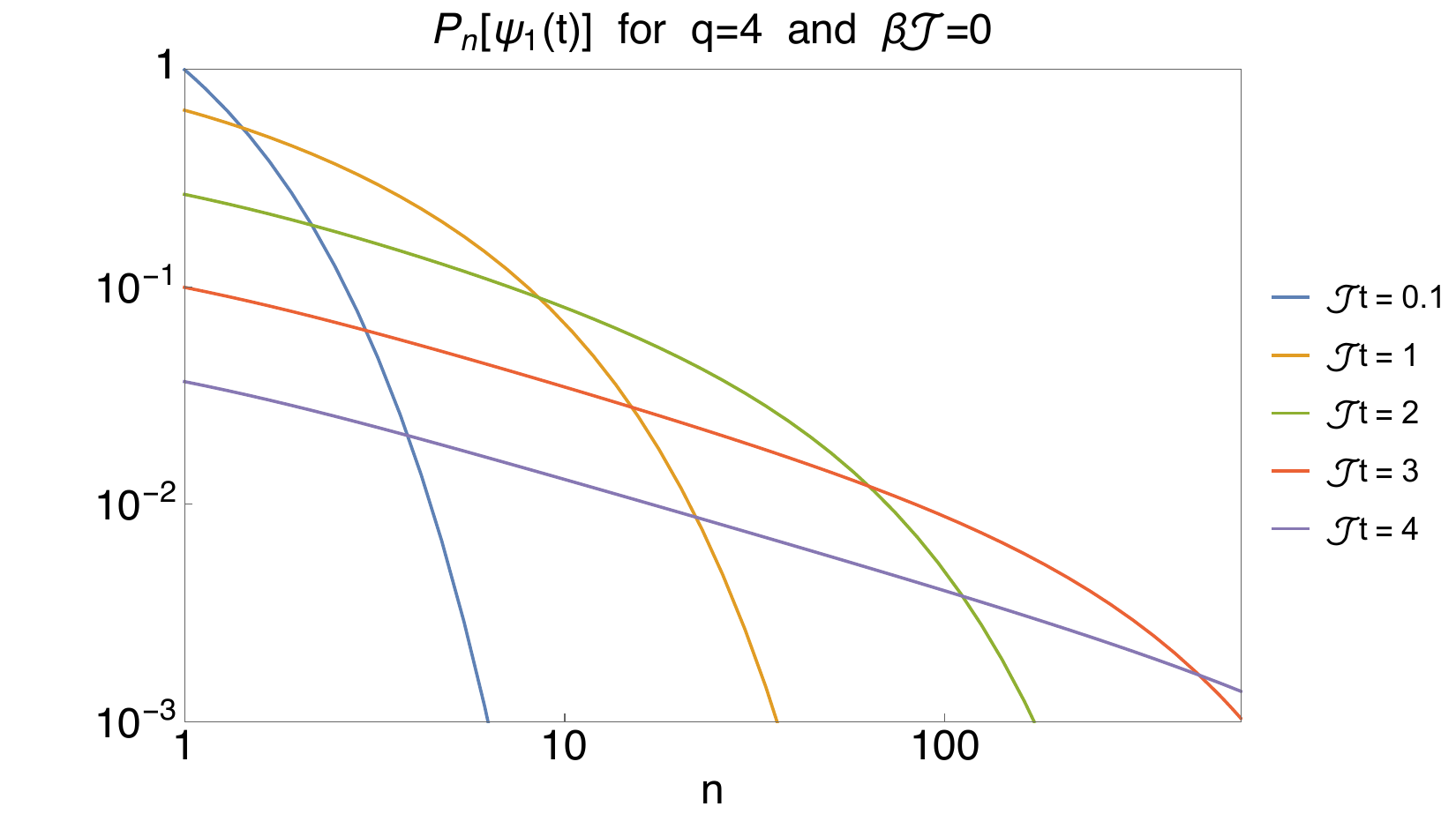} 
\par\end{centering}
\caption{After either dynamical renormalization (\ref{eq:dynamical renormalization})
or time re-parametrization (\ref{eq:time reparameterization}), the
growth distribution $K^{\beta}\left[\psi_{1}\left(t\right)\right]$
takes the same form as the Heisenberg evolution of the operator $\psi_{1}\left(t\right)$
(i.e. the infinite temperature size distribution $P\left[\psi_{1}\left(t\right)\right]$).
This distribution is given by Eq. (\ref{eq:reparametrized growth distribution}),
and we plot it on a log-log scale. Note that it reaches out towards
larger operators exponentially quickly. \label{fig:Growth Distribution}}
\end{figure}
In the Lyapunov regime, we may expand the generating function of the
growth distribution as 
\[
\mathcal{G}_{\mu}\left(\frac{\beta^{+}}{4}+it,\frac{\beta^{-}}{4}+it\right)=e^{-\mu\delta_{\beta}}\sum_{n=0}^{\infty}\binom{-2/q}{n}\left(1-e^{-q\mu\delta_{\beta}}\right)^{n}\left(\frac{\mathcal{J}}{\alpha}\sinh\alpha t\right)^{2n}
\]
where $\delta_{\beta}=\left(\alpha/\mathcal{J}\right)^{2/q}$, and
$\alpha\equiv\alpha_{\mu=0}\left(\beta\mathcal{J}\right)$ is the
smallest positive root of Eq. (\ref{eq:Parameter Constraints}). Grouping
terms by powers of $\exp\left(-\mu\right)$ and using the definition
(\ref{eq:moments of G}), we conclude that the growth distribution
is given by 
\begin{equation}
K_{\delta_{\beta}\left(1+qn\right)}^{\beta}\left[\psi_{1}\left(t\right)\right]=\left(-1\right)^{n}\binom{-2/q}{n}\frac{\left(\frac{\mathcal{J}}{\alpha}\sinh\left(\alpha t\right)\right)^{2n}}{\left(1+\left(\frac{\mathcal{J}}{\alpha}\sinh\left(\alpha t\right)\right)^{2}\right)^{n+\frac{2}{q}}}\label{eq:SYK Growth Distribution}
\end{equation}
where we note that $\left(-1\right)^{n}\binom{-2/q}{n}$ is always
positive for integer $n$. Thus, $K^{\beta}\left[\psi_{1}\left(t\right)\right]\geq0$
and so we have no negative probabilities in the size distribution
of $\psi_{1}\left(t\right)\rho^{1/2}$, since it is given by $P\left[\psi_{1}\left(t\right)\rho^{1/2}\right]=K^{\beta}\left[\psi_{1}\left(t\right)\right]*P\left[\rho^{1/2}\right]$
as shown in section (\ref{subsec:Thermal-Fermion}).

\begin{figure}[tb]
\noindent \begin{centering}
\includegraphics[width=0.8\textwidth]{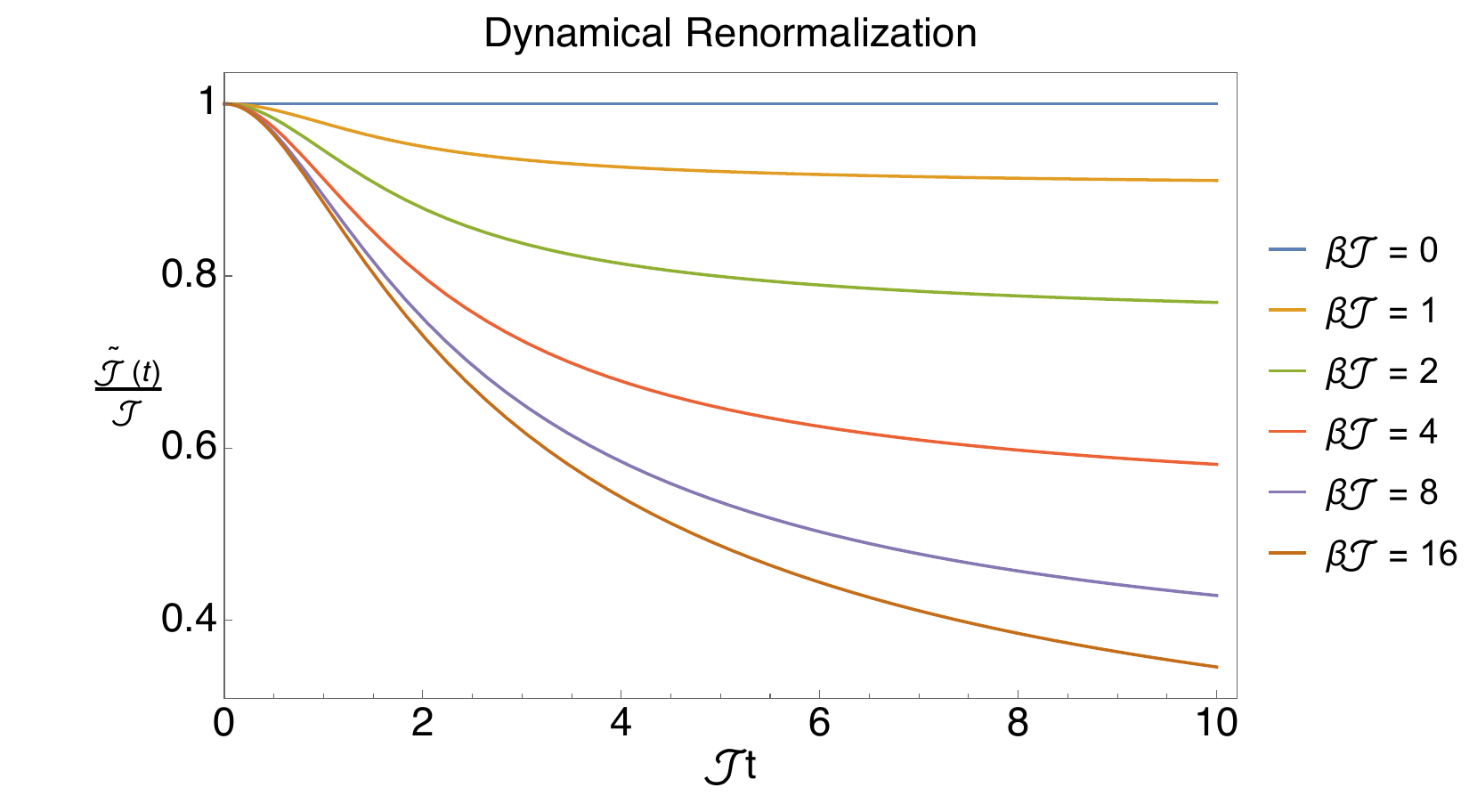} 
\par\end{centering}
\caption{For different values of $\beta\mathcal{J}$, the effective coupling
$\tilde{\mathcal{J}}\left(t\right)$ given by Eq. (\ref{eq:dynamical renormalization})
slows down from $\mathcal{J}$ to $\alpha$ on a timescale of order
$\alpha^{-1}$. As always, $\alpha$ is the smallest root of Eq. (\ref{eq:Parameter Constraints}).
\label{fig:dynrenorm}}
\end{figure}
Interestingly, we see that the growth distribution $K_{\beta}$ (\ref{eq:SYK Growth Distribution})
has a functional form independent of temperature, which we plot in
Fig. (\ref{fig:Growth Distribution}). We can use either of two methods
to expose this phenomenon. One option is to replace the coupling $\mathcal{J}$
with the dynamically renormalized coupling $\tilde{\mathcal{J}}\left(t\right)$
(plotted in Fig. (\ref{fig:dynrenorm})): 
\begin{equation}
\tilde{\mathcal{J}}(t)=\frac{\mathrm{arcsinh}\left(\frac{\mathcal{J}}{\alpha}\sinh\left(\alpha t\right)\right)}{t}=\alpha+\frac{\log\left(\mathcal{J}/\alpha\right)}{t}+\frac{\mathcal{O}\left(e^{-2\alpha t}\right)}{t}\label{eq:dynamical renormalization}
\end{equation}
The other option is to re-parametrize time 
\begin{equation}
\tilde{t}=\frac{1}{\mathcal{J}}\mathrm{arcsinh}\left(\frac{\mathcal{J}}{\alpha}\sinh\left(\alpha t\right)\right)=\frac{\alpha}{\mathcal{J}}t+\frac{\log\left(\mathcal{J}/\alpha\right)}{\mathcal{J}}+\frac{\mathcal{O}\left(e^{-2\alpha t}\right)}{\mathcal{J}}\label{eq:time reparameterization}
\end{equation}
Both methods transform the finite temperature growth distribution
into that of the Heisenberg evolution of the operator $\psi_{1}\left(t\right)$
(i.e. the infinite temperature size distribution $P_{1+qn}\left[\psi_{1}\left(t\right)\right]$)
\cite{Roberts:2018mnp}. For example, using $\tilde{t}$ gives 
\begin{equation}
K_{\delta_{\beta}\left(1+qn\right)}^{\beta}\left[\psi_{1}\left(t\left(\tilde{t}\right)\right)\right]=\left(-1\right)^{n}\binom{-2/q}{n}\frac{\tanh\left(\mathcal{J}\tilde{t}\right)^{2n}}{\cosh\left(\mathcal{J}\tilde{t}\right)^{\frac{4}{q}}}=K_{1+qn}^{\beta=0}\left[\psi_{1}\left(\tilde{t}\right)\right]=P_{1+qn}\left[\psi_{1}\left(\tilde{t}\right)\right]\label{eq:reparametrized growth distribution}
\end{equation}

This temperature-independence is fascinating since the Heisenberg
evolution of $\psi_{1}\left(t\right)$ was obtained in \cite{Roberts:2018mnp}
via fully-dressed Feynman graph calculations. In other words, the
distribution $P\left[\psi_{1}\left(t\right)\right]$ represents the
simple tree graphs such as Fig. \ref{fig:large q diagrams}(a) constructed
using the original SYK Hamiltonian. However, we just showed how the
growth distribution $K^{\beta}\left[\psi_{1}\left(t\right)\right]$
can be easily transformed to $P\left[\psi_{1}\left(t\right)\right]$.
Therefore, since $P\left[\psi_{1}\left(t\right)\rho^{1/2}\right]=K^{\beta}\left[\psi_{1}\left(t\right)\right]*P\left[\rho^{1/2}\right]$
and $P\left[\rho^{1/2}\right]$ is well-peaked, we are led to the
remarkable conclusion the growth dynamics of large-$N$, large-$q$
SYK model is totally universal. In fact, if one waits an initial period
$\alpha^{-1}$ to enter the Lyapunov regime, then one need simply
use the effective size $\delta_{\beta}$ and coupling $\tilde{\mathcal{J}}=\alpha$
for the full growth structure of $\psi_{1}\left(t\right)\rho^{1/2}$
to match that of $\psi_{1}\left(t\right)$.

\subsection{Finite Temperature Epidemic Model}

In this subsection we will discuss the physical interpretation of
the SYK operator growth by relating it to an epidemic model. Intuition
for operator scrambling behavior has been developed by various authors
\cite{Sekino:2008he,Roberts:2018mnp,Aleiner:2016eni,Mueller:1994gb},
resulting in an infection picture for operator growth. An operator
such as $\psi_{1}(t)$ can be expanded in the strings of Majorana
fermion $\Gamma_{I}$. We consider the fermions already included in
the string as ``infected''. Heisenberg evolution of $\Gamma_{I}$
generates a term $\left[\Gamma_{I},H\right]$ which could contain
a few more fermions. For example for SYK model with $q$-body interactions,
in the large $N$ limit most of the terms have one fermion replaced
by $q-1$ other fermions. In order for these $q-1$ fermions to be
``infected'', they must not be already in $\Gamma_{I}$. Therefore
the infection rate depends on the infectable population.

In the simplest infection model for a population of $n_{*}$ individuals,
the rate of infection is proportional to the number of unexposed people
times the number of contagious people 
\begin{equation}
\frac{dn\left(t\right)}{dt}=r\left(1-\frac{n\left(t\right)}{n_{*}}\right)n\left(t\right)\label{eq:SimpleInfection}
\end{equation}
More generally, in various quantum circuit and Hamiltonian systems,
both terms on the right-side of the equation may be raised to various
powers or there may even be a sum of such terms, due to the potential
multi-body nature of the interaction. For example, in SYK, upon a
single commutation with the Hamiltonian, a size $1$ operator becomes
a size $q-1$ operator, so we might expect various powers of $q$
to appear in the above expression. Regardless, in either case sigmoidal
behavior will be produced, which is consistent with general expectations
of four-point functions.

Let us see just how well such a picture can apply to the SYK model.
Taking the derivative of Eq. (\ref{eq:sizegrowthSYK}) and using Eq.
(\ref{eq:Average Growth Scrambles}), we find that during the Lyapunov
regime ($\log N\gg\alpha t\gg1$) 
\begin{align*}
\frac{d}{dt}\left(n\left[\psi_{1}\left(t\right)\rho^{1/2}\right]\right) & \approx\left(2\mathcal{J}\right)\left(1-\frac{n\left[\rho^{1/2}\right]}{n_{*}}\right)^{q/2}\left(n\left[\psi_{1}\left(t\right)\rho^{1/2}\right]-n\left[\rho^{1/2}\right]\right)
\end{align*}
Comparing with the infection equation (\ref{eq:SimpleInfection}),
we have the fundamental rate $r=2\mathcal{J}$ as well one of the
terms being raised to $q/2$ due to the $q$-local nature of the interaction.
However, rather than $\left(1-n\left[\psi\left(t\right)\rho^{1/2}\right]/n_{*}\right)^{q/2}$,
which one may have expected by direct analogy with the infection equation,
we have the static term $\left(1-n\left[\rho^{1/2}\right]/n_{*}\right)^{q/2}=\delta_{\beta}^{q/2}$.
During the Lyapunov regime, these two are the same to leading order
in $N$. Lastly, it appears through the final term that of the large
population $n\left[\psi_{1}\left(t\right)\rho^{1/2}\right]$, only
the small population $n\left[\psi_{1}\left(t\right)\rho^{1/2}\right]-n\left[\rho^{1/2}\right]$
possesses the ability to infect others. Notice that there remains
the large population $n\left[\rho^{1/2}\right]$ who count as having
been exposed, but do not infect others. It is thus natural to view
this group as a vaccinated population.

\begin{figure}[tb]
\begin{centering}
\begin{minipage}[t]{0.45\textwidth}%
\begin{center}
\includegraphics[width=1\textwidth]{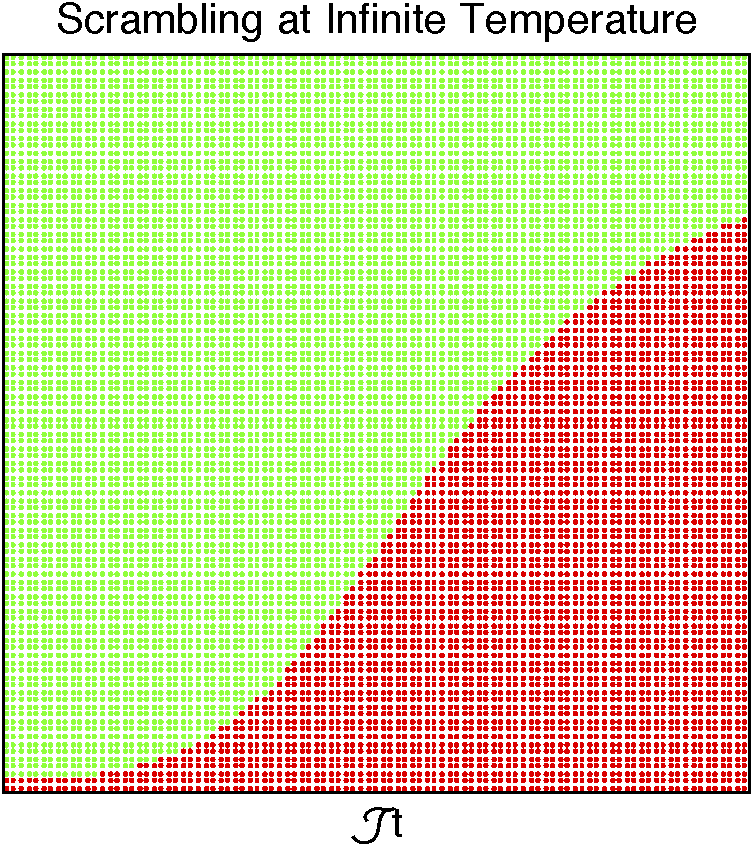} 
\par\end{center}%
\end{minipage}\hfill{}%
\begin{minipage}[t]{0.45\textwidth}%
\begin{center}
\includegraphics[width=1\textwidth]{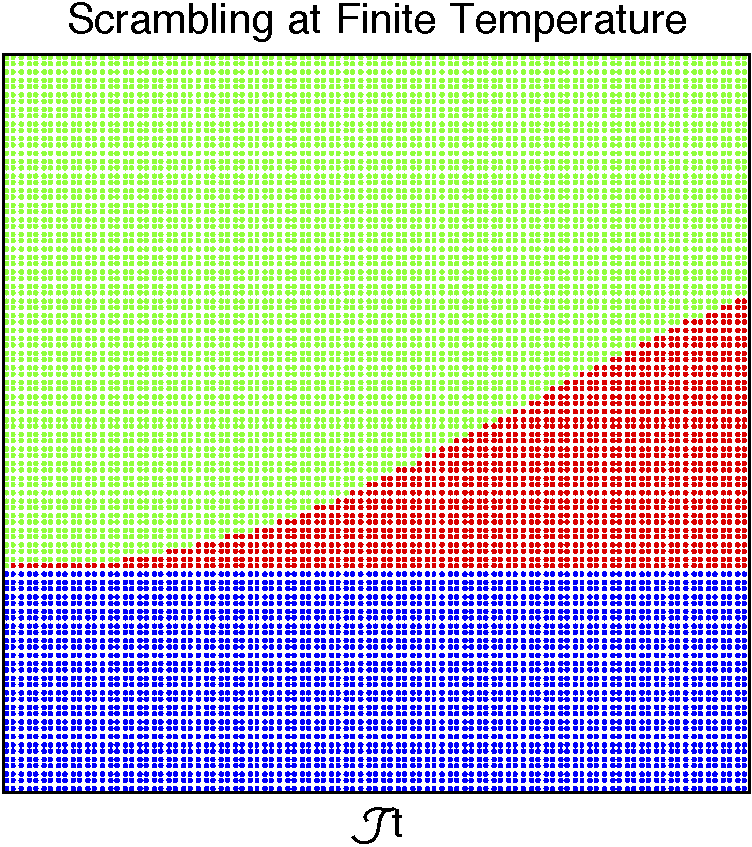} 
\par\end{center}%
\end{minipage}
\par\end{centering}
\caption{Illustration of different epidemics at infinite temperature (left
panel) and finite temperature (right panel). Green dots represents
unexposed individuals, red dots represents contagious individuals,
and blue dots represent vaccinated individuals. Finite temperature
factors such as $\rho^{1/2}$ ``use up'' some of the available flavors
for growth, resulting in collisions like those depicted in Fig. (\ref{fig:collision}).
This effect ends up being well-modeled by an epidemic where these
flavors or individuals count as having been exposed, but do not spread
disease. As a result of this large vaccinated population, it simply
more rare for a contagious individual to encounter an unexposed individual,
even at the start. Hence the rate of infection -- the Lyapunov exponent
-- slows down, as seen in the right figure.\label{fig:epidemics}}
\end{figure}

In other words, after waiting for the dynamical renormalization/time
re-parametrization to settle down, the physics of the four-point function
is well-described by an infection model, with the caveat that only
a small population $n\left[\psi\left(t\right)\rho^{1/2}\right]-n\left[\rho^{1/2}\right]$
possess the ability to infect. In this sense, the operator $\rho^{1/2}$
vaccinates a finite fraction of the $N$ flavors. Now regardless of
whether any particular individual possess the ability to infect, it
remains that a large portion of the population has been exposed, and
thus the probability for any contagious individual to encounter an
unexposed individual is decreased. Consequently, the overall rate
of infection slows down to 
\[
\lambda_{L}=2\mathcal{J}\left(1-\frac{n\left[\rho^{1/2}\right]}{N_{*}}\right)^{q/2}=2\mathcal{J}\delta_{\beta}^{q/2}=2\mathcal{J}\left(G\left(\frac{\beta}{2}\right)\right)^{q/2}=2\alpha
\]
as illustrated in Fig. (\ref{fig:epidemics}).

\section{Discussion\label{sec:discussion}}

The methodology developed in sections (\ref{sec:Operator-Distributions-and})
and (\ref{sec:Thermal-Operators}) is very powerful, as it applies
to all fermionic systems. Specifically, determining the system's full
growth distribution amounts to calculating the twisted (\ref{eq:Two-Point Bdry Conds})
two-point function $\mathcal{G}_{\mu}$ followed by inverse transforming
in $\mu$ (\ref{eq:moments of G}). The large-$N$ saddle point technique
and the large-$q$ simplification enabled us to obtain a closed solution
in SYK. Even if analytics are too difficult, this analysis can be
effectively implemented numerically for many classes of models.

These techniques also allowed us to compute a four-point function,
since Eq. (\ref{eq:4ptfiniteT}) shows that the average size of the
operator $\psi\left(t\right)\rho^{1/2}$ (\ref{eq:sizegrowthSYK})
gives the value of a certain four-point function. We can generalize
and calculate arbitrary four-point functions by moving the twist (\ref{eq:Two-Point Bdry Conds})
to other locations. This has the non-trivial consequence that the
twisted two-point function solves the ladder kernel\cite{Maldacena:2016hyu,KitaevTalks}.
In practice solving the former can be substantially easier than solving
the latter. As an example, in \cite{Streicher:2018chaos2} we use
this ``twisted'' technique to derive an elegant expression for the
large-$q$ SYK four-point function at arbitrary coupling and temperature.
Like the growth distribution methodology, this new method for calculating
four-point functions works for all fermionic systems.

The dynamical renormalization of the coupling (\ref{eq:dynamical renormalization})
plays a central role in this work. It will be important to understand
this in a deeper and more general context. The success of the modified
infection model in capturing the thermal operator growth suggests
that the principle underlying the finite temperature slowdown in SYK
is competition for Majorana flavors. The presence of various powers
of the thermal state $\exp\left(-\beta H\right)$ ``uses up'' some
finite fraction of the flavors. Consequently, when we apply a single
fermion, there is a fractional probability for it to become absorbed
and thus its size is renormalized (\ref{eq:average size increase})
to a value based upon the percentage $\delta_{\beta}$ of ``unused''
flavors. Now, the renormalized coupling $\mathcal{J}\left(t\right)$
(\ref{eq:dynamical renormalization}) slows down during time-evolution.
We believe that this occurs due to the same principle, but have not
yet fully understood the mechanism. Our belief is motivated by the
empirical observation that the Lyapunov exponent is a power of the
percentage $\delta_{\beta}$ of ``unused'' flavors
\[
\lambda_{L}=\lim_{t\gg t_{dissipation}}2\mathcal{J}\left(t\right)=2\mathcal{J}\left(\delta_{\beta}\right)^{q/2}\equiv2\mathcal{J}\left(1-\frac{n\left[\rho^{1/2}\right]}{n_{*}}\right)^{q/2}
\]

This kind of sigmoidal operator growth is generic in many-body chaos.
However, without Majorana fermions, the manner in which the thermal
factors interfere with operator growth must be more complicated, as
there is not a bit-like notion of ``using up'' a flavor. Sigmoidal
behavior signals the existence of a competition for some finite resource.
For SYK, this resource was flavor; we only have $N$ flavors with
which to grow operators, so eventually we will be led to flavor collisions
as in Fig. (\ref{fig:collision}). However, flavor competition is
only one aspect of competition for a more general resource. The question
remains: what do operators compete for during evolution? Is there
some sort of ``operator entropy''? Perhaps when summed across ``all
operators'' at finite temperature, there is always a fixed amount
of total correlation with an initial simple operator due to unitarity.
A better understanding of such a resource would give an organization
to operator dynamics at different energy scales.

Our results have interesting implications for the holographic dual
of the SYK model. This is simplest to understand when we explicitly
express our results in terms of the doubled theory. We defined an
entangled orthonormal basis for the doubled theory using the eigenstates
(\ref{eq:orthonormal basis states}) of the size operator (\ref{eq:size operator}).
Taking the state $\psi_{1}^{L}\left(t\right)\ket{TFD}$\footnote{If we replace $t\rightarrow-t$, then this is the precursor state
$\psi_{1}^{L}\left(-t\right)\ket{TFD}$, where the ``boundary''
operator $\psi_{1}^{L}$ acted upon the thermofield double state at
time $-t$ \cite{Stanford:2014jda,Roberts:2014ifa,Roberts:2014isa,Shenker:2014cwa}.}, we related its size wave-function squared (i.e. $P_{n}\left[\psi\left(t\right)\rho^{1/2}\right]\equiv\left|\mel{n}{\psi_{1}^{L}\left(t\right)}{TFD}\right|^{2}$)
to the size wave-function squared for the thermofield double state
(i.e. $P_{n}\left[\rho^{1/2}\right]\equiv\left|\braket{m}{TFD}\right|^{2}$)
\[
\left|\mel{n}{\psi_{1}^{L}\left(t\right)}{TFD}\right|^{2}=\sum_{m=0}^{n}K_{n-m}^{\beta}\left[\psi_{1}(t)\right]\left|\braket{m}{TFD}\right|^{2}
\]
isolating the time-dependence into the growth distribution $K^{\beta}\left[\psi_{1}\left(t\right)\right]$
(\ref{eq:convolution equation}). Using this, we found the ``average
growth'' of $\psi_{1}\left(t\right)$ (\ref{eq:Average Growth})
at low temperatures to be $1+2\left(\beta\mathcal{J}\sinh\left(\pi t/\beta\right)/\pi\right)^{2}$,
which was shown in \cite{Brown:2018kvn} to exactly match the classical
momentum dynamics of a ``boundary'' particle falling into a near-extremal
black hole. That is, the average growth of an SYK fermion exactly
matches the average momentum of an infalling particle in a $NAdS_{2}$
black hole.

It is a striking result of our analysis that the \emph{full }size
wavefunction squared of the SYK fermion precisely relates to the \emph{full}
momentum wavefunction squared of the infalling particle. The universal
form (\ref{eq:reparametrized growth distribution}) of the growth
distribution $K^{\beta}\left[\psi_{1}\left(t\right)\right]$ precisely
gives the squared coefficients of the $AdS_{2}$ momentum bulk-to-boundary
propagator. Exploring this connection will be an important focus of
future work\footnote{One next step will be to perform this analysis for a different geometry,
and a natural place to do so is in the set-up created by \cite{Maldacena:2018lmt}.
It was found that adding the scrambling distance operator (\ref{eq:FracScramDisOp})
to the doubled SYK Hamiltonian causes the low-energy limit to eventually
cross a Hawking-Page transition, forming a global $NAdS_{2}$ geometry
instead of a $NAdS_{2}$ black hole.}.

\section*{Acknowledgements}

We are grateful to Adam Brown, Tarun Grover, Yingfei Gu, Guy Gur-Ari,
Matt Hastings, Andy Lucas, Daniel Ranard, Dan Roberts, Phil Saad,
Steve Shenker, Eva Silverstein, Douglas Stanford, Lenny Susskind,
Brian Swingle, Aron Wall, and Ying Zhao for extremely useful discussions.
This work is supported by the National Science Foundation grant 1720504
(XLQ), and the Simons Foundation (XLQ and AS). This work is supported in part by the U. S. Department of Energy award DE-SC0019380 (XLQ).

\bibliographystyle{utphys}
\bibliography{draft}

\end{document}